\documentclass[11pt]{article}
\usepackage[utf8]{inputenc}
\usepackage{natbib}
\usepackage{amsfonts}
\usepackage{amsmath}
\usepackage{amssymb}
\usepackage{amsthm}
\usepackage{graphicx}
\usepackage{subcaption}
\usepackage{url}
\usepackage{arydshln}
\usepackage{setspace}
\usepackage{changebar}
\usepackage{xcolor}

\newcommand{\Ne}{{\mathcal{N}}}
\newcommand{\calF}{{\mathcal{F}}}
\newcommand{\Bin}{\operatorname{Bin}}
\newcommand{\Poi}{\operatorname{Poi}}
\newcommand{\Var}{\operatorname{Var}}
\newcommand{\Cov}{\operatorname{Cov}}
\newcommand{\cor}{\operatorname{Cor}}
\newcommand{\MVN}{\operatorname{MVN}}
\newcommand{\relu}{\operatorname{relu}}
\newcommand{\arcsinh}{\operatorname{arcsinh}}
\newcommand{\softplus}{\operatorname{softplus}}
\newcommand{\bracenom}{\genfrac{\lbrace}{\rbrace}{0pt}{}}

\DeclareMathOperator*{\argmin}{arg\,min}

\newcommand{\diag}{\operatorname{diag}}

\newcommand{\reals}{{\mathbb{R}}}
\newcommand{\nats}{{\mathbb{N}}}
\newcommand{\natsN}{\nats_{0}}
\newcommand{\E}{{\mathbb{E}}}
\newcommand{\bX}{\mathbf{X}}

\newtheorem{theorem}{Theorem}[section]

\newtheorem{lemma}[theorem]{Lemma}
\newtheorem{remark}{Remark}
\newtheorem{proposition}{Proposition}
\newtheorem{definition}{Definition}

\newcommand{\calA}{\mathcal{A}}

\title{New Methods for Network Count Time Series}
\author{Hengxu Liu and Guy Nason\thanks{Corresponding
            author: {\tt gnason@imperial.ac.uk}. ORCID id: 0000-0002-4664-3154}\\
    Imperial College London\thanks{Department of Mathematics,
        Huxley Building, Imperial College London,
        180 Queen's Gate, London SW7 2AZ.}}
\date{4 December 2023}

\begin{document}

\doublespacing

\maketitle

\begin{abstract}
   The original generalized network autoregressive models are poor for modelling count data as they are based on the additive and constant noise assumptions, which is usually inappropriate for count data. We introduce two new models (GNARI and NGNAR) for count network time series by adapting and extending existing count-valued time series models. We present results on the statistical and asymptotic properties of our new models and their estimates obtained by conditional least squares and maximum likelihood. We conduct two simulation studies that verify successful parameter estimation for both models and conduct a further study that shows, for negative network parameters, that our NGNAR model outperforms existing models and our other GNARI model in terms of predictive performance. We model a network time series constructed from COVID-positive counts for counties in New York State during 2020--22 and show that our new models perform considerably better than existing methods for this problem.
\end{abstract}


\section{Introduction}

\subsection{Network time series}
A network time series is
the pair $\left[ \{X_t\}_t, G \right]$, where $G = (V, E)$ is a graph (or network), where
$V$ is a set of vertices or nodes, with $|V| = N$,
and edge set $E$
and $\{X_t\}_t$ is a \mbox{$N$-dimensional} multivariate time series $X_{i, t}$ for $i \in V$ and times $t=1, \ldots, T$
for some integer $T>0$.
We use the notation $i \leftrightsquigarrow j$,
if $i\in V$ is directly connected to $j \in V$.
Given a subset of nodes $A \subset V$, then the
neighbourhood set of $A$ is defined by
\begin{equation}
\Ne(A) = \{ j \in V : j\leftrightsquigarrow i,
    i \in A\}.
\end{equation}
and further define the $r$th stage neighbours
for $r>1$ by
\begin{equation}
\Ne^{(r)} (i) = \Ne \{ \Ne^{(r-1)} (i) \} / \cup_{q=1}^{r-1} \Ne^{(q)} (i).
\end{equation}
For example, $\Ne^{(2)}(i)$ are the neighbours
of the immediate neighbours of vertex $i$,
not including those immediate neighbours or
$i$. This article focuses on the situation
where the $X_{i, t}$ multivariate time series
are counts, that is integers greater than
or equal to zero.

\subsection{GNAR models}

A particular recent network time series
model is the generalized network autoregressive model (GNAR) introduced by
\cite{gnar16}, see also \cite{nvar} and \cite{gnar}. The GNAR model assumes that each node
(variable) of the multivariate
time series is influenced by an standard autoregressive term and contributions from neighbours in the network
at earlier times. A GNAR$(p, [s_1, \ldots, s_p])$ model is given by:
 \begin{equation}
 \label{eq:narps}
X_{i, t} = \sum_{j=1}^p \left(  \alpha_{i,j} X_{i, t-j} + \sum_{c=1}^C \sum_{r=1}^{s_j}
\beta_{j, r, c} \sum_{q \in \Ne^{(r)}(i)}
	\omega_{i, q, c}^{(t)} X_{q, t-j} 
	\right)
	+ \epsilon_{i, t},
 \end{equation}
 for $i=1, \ldots, N; t=p+1, \ldots, T$ and
 where $\{ \epsilon_{i, t} \}$ are a set of mutually uncorrelated random variables with
 mean zero and variance of $\sigma^2$.

 The components are the GNAR model
 in~\eqref{eq:narps} are 1.\ the
 autoregressive parameters,
 $\{\alpha_{i, j}\}_{i=1, \ldots, N; j=1, \ldots p}$
 explain how past values of $X_{i, \cdot}$
 contribute to $X_{i, t}$. Every
 vertex (or node) in the graph
 has their own autoregressive sequence.
 For some data sets, it is possible
 to set $\alpha_{i, j} = \alpha_j$. In
 other words, a common
 $\{ \alpha_j\}_{j=1, \ldots, p}$ applies
 to each vertex and the model is then
 called a {\em global} GNAR process.
 2.\ The term $\sum_{q \in \Ne^{(r)}(i)}
	\omega_{i, q, c}^{(t)} X_{q, t-j}$
 elicits a contribution from
 each $r$th stage neighbour, $q$, of
  vertex $i$, lagged by time $j$ relating
  to covariate type $c$. The
  $w^{(t)}_{i, q, c}$ are called
  connection weights, which are often
  related to the local positioning of
  neighbours of vertex $i$ relative to
  $i$ and each other. The weights could be inverse distance
  weights, see~\cite{gnar} for a definition and an example of how they 
  are used.
  3. The $s_j$ control the maximum number of
  stages of neighbours at lag $j$ for node $i$.

\subsection{Features of GNAR models}
Once formulated and model selection completed,
GNAR models can be fit efficiently 
and rapidly by using existing linear modelling
software. \cite{gnar} and \cite{nw20} demonstrate that well-fitting GNAR-based models 
are highly parsimonious and also deal with missing data well.

However, the in their standard form GNAR models are
auto- and cross-regressive models where $X_{i, t} \in \reals$ and,
hence, unsuitable for modelling
count data, especially when the counts are low. For example, fitting regular GNAR
models to low-count data can result in undesirable negative forecasts for future $X_{i, t}$.
More subtly, the base GNAR models
assume constant unconditional variance,
unrelated to the mean whereas for, e.g.\ Poisson
count models we would require the variance
to be related to the mean.

\subsection{Some univariate count time series models}
\label{sec:review}
We briefly review three types of popular time series
models: the INAR, GAR and NAR models and some
closely related ones. See \cite{cts21} for a comprehensive recent review.

The {\em INteger-valued AutoRegressive} model
is based on thinning operations. 
The INAR$(1)$ model was introduced by
\cite{inar1} and the general INAR$(p)$
model by \cite{inarp}. The 
binomial thinning operation
is defined as follows. Let $Y_1, Y_2, \ldots$
be a collection of independent and
identically distributed Bernoulli
random variables with (probability of 
success) parameter $q \in (0, 1)$. Let
$X$ be a non-negative integer random
variable. The binomially-thinned random
variable $Y$ is given by
\begin{equation}
    Y = q \circ X = \sum_{i=1}^X Y_i.
\end{equation}
Clearly, this implies that $0 \leq Y \leq X$,
$Y|X \sim \Bin(X, q)$ and so 
\begin{equation}
\label{eq:condmeanvar}
\E (Y| X) = qX \text{ and } \Var(Y|X) = q(1-q)X.  
\end{equation}
The marginal distribution of $Y$ depends
on $X$. So, for example, if $X \sim \Poi(\lambda)$,
then $Y|X \sim \Poi(\lambda q)$,
where $\Poi$ is the Poisson distribution.

Let $\{ \alpha_j \in (0,1)\}_{j=1}^p$, then
the univariate INAR$(p)$ model for count time series, $X_t$
is given by
\begin{equation}
X_t = \sum_{j=1}^p \alpha_j \circ X_{t-j} + \epsilon_t.
\end{equation}
Here $X_t \in \natsN$, the set of non-negative
integers, and $\epsilon_t \in \natsN$ is
a set of uncorrelated random variables.
If the $\{ \epsilon_t \}$ are Poisson-distributed,
then $X_t$ is then called a Poisson-INAR$(p)$
process.

INAR$(p$) processes share many similarities with  AR$(p)$
processes, e.g.\ the autocorrelation structure~\citep{inarp} and
conditions for stationarity. However, there are differences
too, such as with the conditional variance or the precise form
of the autocorrelation function. Indeed, the autocorrelation
of an INAR$(p)$ process has the same form as an 
ARMA$(p, p-1)$ process, see~\cite{inarp-2}.

For INAR processes $\cor(X_t, X_{t-j}) = \alpha_j$ for
$j=1, \ldots, p$, and since $\alpha_j \in [0, 1]$, this means
that INAR processes do not admit negative
autocorrelations, which obviously
means that they are not good models for data that exhibit
such negative autocorrelations.

The generalized autoregressive  GAR$(p)$ model, without covariates, is characterised by
the following `mean-relation'
\begin{equation}
    g(\mu_t) = \sum_{j=1}^p \alpha_j \calA(X_{t-j}),
\end{equation}
where $\mu_t = \E (X_t)$, $g$ is a link function,
$\alpha_j \in \reals$ and $\calA$ is some 
function that modifies the autoregressive relations.
GAR processes are related to generalised linear models.
GAR$(p)$ models are special cases of the 
GARMA$(p, q)$ model as introduced by~\cite{garma}.
Often, if $X_t \sim \text{GARMA}(p,q)$ given past history follows
some exponential family, and popular choices for the conditional
distribution are
Poisson, binomial and gamma. Many
popular nonlinear models extended from the 
integer-valued generalized autoregressive
conditional heteroskedasticity (INGARCH) model
use the same link function idea as GARMA. For example,
the log-linear Poisson autoregression from
\cite{loglinear} or the \mbox{$\softplus$-INGARCH} model
from~\cite{softplusgnarch}.

The {\em nonlinear autoregressive} (NAR)
process~\cite{nar} has similarities to GAR ---
they both involve a link or response function as
follows:
\begin{equation}
    X_t = \lambda \left( \sum_{j=1}^p X_{t-j} \right)
    + \epsilon_t,
\end{equation}
where $\lambda$ is the response function and
$\epsilon_t$ are i.i.d.\ random variables.
In some, but not all, cases it is possible
to convert a GAR process into a NAR process.

\subsection{Poisson network autoregression}
The first network time series model for count data was
the Poisson network autoregression (PNAR) and Poisson GNAR, which are a count-valued
network time series models introduced by~\cite{pnar} based
on the network autoregression models from~\cite{nvar} and~\cite{gnar16,gnar}, respectively.
The linear PNAR$(p)$ model assumes for vertex\
$i$ and time $t$ that
$X_{i, t}\sim \Poi(\lambda_{i, t})$,
where
\begin{equation}
\label{eq:pnarlinpred}
\lambda_{i, t} = \beta_0 +
    \sum_{m=1}^p \beta_m n_i^{-1} \sum_{j=1}^N a_{i, j}
        X_{j, t-m} + \sum_{m=1}^p \alpha_m X_{i, t-m},
\end{equation}
where $\beta_0, \beta_m, \alpha_m$ are non-negative
for $m=1, \ldots, p$ are network influence and
autoregression parameters respectively,
$n_i$ is the out-degree of node $i$,
and $A = (a_{i, j})_{i, j}^{N^2}$ is the adjacency matrix
of a graph $G$.

The PNAR model permits interdependence among nodes at
time $t$ and this correlation is induced via copula methods, which
depends on unknown parameters in addition to the $\alpha$s and
$\beta$s above. Interesting conditions for stationarity and
ergodicity for PNAR$(p)$ are
\begin{equation}
    \rho \left( \sum_{m=1}^p G_m \right) < 1,
\end{equation}
where $G_m = \beta_m W + \alpha_m I_N$,
$W = \diag(n_1^{-1}, \ldots, n_N^{-1})A$ and $\rho$ is the
spectral radius of a matrix.

\cite{pnar} further present a count data extension of the
GNAR model~\cite{gnar16,gnar} termed the Poisson GNAR
where the conditional mean is based on GNAR components as
\begin{equation}
\lambda_{i, t} = \alpha_0 + \sum_{j=1}^p (\alpha_{i, j} X_{i, t-j}
 + \sum_{r=1}^{s_j} \beta_{j, r} \sum_{q \in \Ne_t^{(r)}(i)}
 w_{i, q}^{(t)} X_{q, t-j} ),
\end{equation}
where the $\alpha_0, \alpha_{i, j}, \beta_{j, r}$ are nonnegative.
The PNAR model is a special case of the Poisson GNAR model.

\cite{pnar} also consider a log-linear PNAR model
$X_{i, t}\sim \Poi \{ \exp(\nu_{i, t}) \}$ where the linear
predictor, $\nu_{i, t}$, is identical to~\eqref{eq:pnarlinpred}
except that the $X$ terms are replaced by $\log(X+1)$ and
the $\alpha$ and $\beta$ parameters can be real numbers.
The $+1$ in the $\log$ term in the linear predictor is to handle
zero values of $X$. Parameter estimation for both the linear
and log-linear PNAR models is performed by
quasi-maximised likelihood estimation (QMLE).

\section{Count Network Time Series: Two New Models}
This section introduces two new models for network count series: the
generalized network autoregressive integer-valued (GNARI)
model, which is adapted from INAR models and the
nonlinear generalized network autoregressive (NGNAR) model, which is
adapted from GAR models.

Proofs of all results are contained in the appendix.

\subsection{The GNARI model}
As with INAR, GNARI processes
replace multiplications by thinning.

\subsubsection{GNARI model definition}
The  GNARI$(p, [s_1, \ldots, s_p])$ process is defined by
\begin{equation}
\label{eq:gnari}
    X_{i, t} = \sum_{j=1}^p \{ \alpha_{i, j} \circ X_{i, t-j} +
    \sum_{r=1}^{s_j} \beta_{j, r} \circ 
    \sum_{q \in \Ne_t^{(r)}(i) } w_{i, q}^{(t)} \circ X_{q, t-j} \}
    + \epsilon_{i, t},
\end{equation}
where $\circ$ denotes thinning as before,
$\alpha_{i, j}, \beta_{j, r} \in [0, 1]$,
$\epsilon_{i, t}$ is assumed to be non-negative independently
distributed for each node $i$ and time $t$, and identically
distributed for the same $i$, i.e.\ for each $i, t$
we have $\E (\epsilon_{i, t}) = \lambda_i$ for some
$\lambda_i \in \reals^{+}$.  This specification permits
us to assign a different mean and variance for each node $i$,
but we can choose the make the process `global', i.e.\
$\lambda_i=\lambda$ as with the `global-$\alpha$ specification
of the original GNAR processes. 
We  assume the $\epsilon_{i, t}$
are Poisson-distributed.

GNARI models share the same limitation of not
permitting negative correlations as INAR. However,
they are a popular model in the regular time series
case and worth study. Parameter estimation
can be carried out by conditional least-squares, which we
develop next.

\subsubsection{GNARI conditional distribution and stationarity}
The conditional distribution of $X_{i, t} | \calF_{t-1}$,
where $\calF_{t-1}$ is the \mbox{$\sigma$-algebra} generated
by $\bX_{t-1}, \bX_{t-2}, \ldots$ can be accessed
via  moment generating functions (MGFs). We now drop the filtration notation and
the $t$ from the connection weights, i.e.\ $w^{(t)}_{i, q}$
just becomes $w_{i, q}$ and all distributions are 
conditioned on the history $\calF_{t-1}$.

We  introduce the additional notation
$Y^{(r)}_{i, j, t}$ to be the contribution of the $r$th-stage
neighbours of node $i$ that are $j$ time steps
prior to time $t$, i.e.\ 
\begin{equation}
    Y^{(r)}_{i, j, t} = \sum_{q \in \Ne_i^{(r)}(i)} 
    w_{i, q} \circ X_{q, t-j},
\end{equation}
which is the second part of the second term of equation~\eqref{eq:gnari}.
For definiteness let the elements of
$\Ne^{(r)}_t (i)$ be $q_1, \ldots, q_m$, for some
integer $m$ (these are the $r$th-stage neighbours
of node $i$). By construction
$Y_{i, j, t}^{(r)}$ has a Poisson binomial distribution
with parameters
\begin{displaymath}
w_{i, q_1}, \ldots, w_{i, q_1},
w_{i, q_2}, \ldots, w_{i, q_2}, \ldots,
w_{i, q_m}, \ldots, w_{i, q_m},
\end{displaymath}
where
each $w_{i, q_\ell}$ is repeated $X_{q_{\ell}, t-j}$ times.

Now, let
\begin{equation}
    Z_{i,j, t}^{(r)} = \beta_{j, r} \circ Y^{(r)}_{i, j, t}
    = \sum_{k=1}^{Y_{i, j, t}^{(r)}} B_{j, r, k},
\end{equation}
where $B_{j, r, k}$ are Bernoulli$(\beta_{j, r})$
random variables.
The $Z_{i, j, t}^{(r)}$ quantity encapsulates the full second term in~\eqref{eq:gnari}.
\begin{lemma}\label{lem:Z}
The distribution of $Z_{i, j, t}^{(r)}$ is Poisson binomial with parameters
\begin{displaymath}
\beta_{j, r} w_{i, q_1}, \ldots, \beta_{j, r} w_{i, q_1}, \ldots, \beta_{j, r} w_{i, q_m}, \ldots, \beta_{j, r}w_{i,q_m},
\end{displaymath}
where $\beta_{j, r} w_{i, q_\ell}$ is repeated $X_{q_{\ell}, t-j}$ times.
\end{lemma}
Returning to the GNARI model~\eqref{eq:gnari} for a moment, if the $\epsilon_{i, t}$ are Poisson distributed with
constant mean $\lambda$, then the conditional distribution of $X_{i, t}$ will be the sum of Poisson
binomial distributions and a Poisson distribution, for which we believe there is no closed form.
From the previous result, one can see that numerical
approximations for the distribution are feasible for computation of conditional maximum likelihood,
but would be highly computationally intensive.

The conditional variance for $X_{i, t}$ given $\mathcal{F}_{t-1}$ is 
\begin{align}
    \Var (X_{i,t}|\mathcal{F}_{t-1})  &  =  \lambda_i + \sum_{j=1}^p \{ \alpha_{i,j} (1-\alpha_{i,j}) X_{i,t-j}
    \label{gnari_cond_var}\\
        & + \sum_{r=1}^{s_j}\sum_{q \in \mathcal{N}^{(r)}_t(i)} \beta_{j,r}w_{i,q} (1-\beta_{j,r}w_{i,q}) X_{q,t-j}\},
\end{align}
using~\eqref{eq:condmeanvar}.
Hence a large conditional mean will cause a large conditional variance.
This observation aligns
GNARI processes much 
more to count data processes
for which the variance
is strongly related to the mean,
unlike standard GNAR where they
are separate.
\begin{remark}
Another possible variant of the GNARI process($p,[s_1,\dots,s_p]$) is

\begin{equation}\label{eq:gnari_variant}
    X_{i,t} = \sum_{j=1}^{p} \{\alpha_{i,j} \circ X_{i, t-j} +\sum_{r=1}^{s_j} \sum_{q\in \mathcal{N}^{(r)}_t(i)} (\beta_{j,r}w_{i,q}^{(t)} \circ X_{q,t-j})\} + \epsilon_{i,t}.
\end{equation}

By definition we have that $\sum_{q\in \mathcal{N}^{(r)}_t(i)} (\beta_{j,r}w_{i,q}^{(t)} \circ X_{q,t-j})\}|\mathcal{F}_{t-1}$ has a Poisson binomial distribution with parameters
\begin{equation}
\beta_{j,r}w_{i,q_1}, \ldots, \beta_{j,r}w_{i,q_1},\ldots, \beta_{j,r}w_{i,q_m},\ldots,\beta_{j,r}w_{i,q_m},
\end{equation}
where $\beta_{j,r}w_{i,q_l}$ is repeated $X_{q_l,t-j}$ times, which is the same as that of $Z_{i,j}^{(r)}$. It thus follows that the two processes \eqref{eq:gnari} and \eqref{eq:gnari_variant} have the same conditional distribution and thus equal in distribution for any same initial distribution. This variant will be useful later to establish  stationarity conditions for the GNARI process.
\end{remark}
We next examine parameter conditions for second-order stationarity.
\begin{lemma}\label{lem:sty}
A sufficient condition for the GNARI($p,[s_1,\dots,s_p]$) to have a unique stationary solution is that the parameters
satisfy the following inequality.
\begin{equation} \label{eq:gnari_stat}
    \sum_{j=1}^p(|\alpha_{i,j}|+\sum_{r=1}^{s_r} |\beta_{j,r}|) < 1, \ \ \  \forall i =1,\dots,N
\end{equation}
\end{lemma}

%
%

\subsubsection{GNARI process autocovariance}
We now derive the autocovariance function $\Gamma(h) = \Cov(\mathbf{X}_t, \mathbf{X}_{t-h})$
for~\eqref{eq:gnari_variant}, under stationarity.
From the proof of Lemma~\ref{lem:sty},
a GNARI($p,[s_1,\dots,s_p]$) process is equivalent to the MGINAR(1) process as defined in \eqref{eq:mginar1}. Then, by~\cite{mginar} Section~4,
the autocovariance function for~\eqref{eq:mginar1}, $\Gamma'(h)$, satisfies 
\begin{equation}
    \Gamma'(h) = \begin{cases}
    A \Gamma'(1)^T + \diag(B\mu_Y) + \Sigma_e, & h=0,\\
    A^h \Gamma'(0), & h \geq 1,
    \end{cases}
\end{equation}
where $B$ is the variance matrix corresponding to the thinning operation $A \circ \cdot$, $\mu_Y = \mathbb{E}[\mathbf{Y}_t]$, and $\Sigma_e = \Var[\mathbf{e}_t]$ and $\mathbf{Y}_t$ is defined in~\eqref{eq:mginar1}. 
For GNARI, let $*$ be entry-wise multiplication.
Thus, we have
\begin{equation}
    B = \begin{bmatrix}
A_1*(1-A_1) & A_2*(1-A_2) & \dots &A_{p-1}*(1-A_{p-1})&A_p*(1-A_p)\\
0 & 0 & \dots &0 &0\\
\vdots &\vdots &\ddots&\vdots&\vdots\\
0&0&\dots&0&0
\end{bmatrix},
\end{equation}
\begin{equation}\label{eq:mu_Y}
    \mu_Y  = (I-A)^{-1}(\lambda_1, \dots, \lambda_N, 0, \dots, 0)^T
\end{equation}
and $\Sigma_e = \diag (\lambda_1, \dots, \lambda_N,0, \dots, 0)$. For instance, we have
\begin{equation}\label{eq:Gamma_0}
    \Gamma'(0) = (I-A)^{-1}\{\text{diag}(B\mu_Y) + \Sigma_e\}(I-A^T)^{-1},
\end{equation}
which exists under stationarity.

It is easy to show that the autocovariance function for~\eqref{eq:gnari_variant}, $\Gamma(h+j)$, is the $[jN+1:(j+1)N, jN+1:(j+1)N]$ submatrix of $\Gamma'(h)$. More specifically, the autocovariance function $\Gamma'(h)$ can be written in terms of $\Gamma(h+j)$ as
\begin{equation}\label{autocov}
    \Gamma'(h) = \begin{bmatrix}
    \Gamma(h) & \Gamma(h+1) & \dots &\Gamma(h+p-1)\\
    \Gamma(h+1) & \Gamma(h) & \dots &\Gamma(h+p-2)\\
    \vdots &\vdots &\ddots&\vdots\\
    \Gamma(h+p-1) & \Gamma(h+p-2) & \dots &\Gamma(h)
    \end{bmatrix},
\end{equation}
from which we can obtain any specific autocovariance function $\Gamma(h)$ of interest.

\subsubsection{Conditional least squares estimation}
\label{sec:evcls}
For ease of notation, we consider only the global $\alpha$ and global $\lambda$ case.
The local $\alpha$ and local $\lambda$ case is a straightforward generalization.

Let $\mathbf{\theta} = (\alpha_{1}, \beta_{1,1}, \ldots, \beta_{1,s_1}, \ldots, \alpha_{p}, \beta_{p,1}, \ldots, \beta_{p,s_p}, \lambda)^T$ be the parameter of interest, $\mathcal{F}_t$ be the $\sigma$-algebra generated by $\mathbf{X}_{t},\mathbf{X}_{t-1}, \ldots$, then the conditional least squares estimator $\hat{\mathbf{\theta}}^{(n)} = \argmin_{\theta} Q_n(\theta)$ minimizes
\begin{align}
 Q_n(\mathbf{\theta}) &= \sum_{t=1}^n ||\mathbf{X}_t - \mathbb{E}_\theta (\mathbf{X}_t|\mathcal{F}_{t-1})||^2\\
 &= \sum_{t=p+1}^n \sum_{i=1}^N[X_{i,t} - \sum_{j=1}^{p} \{\alpha_{j}  X_{i, t-j} +\sum_{r=1}^{s_j} \beta_{j,r}  \sum_{q\in \mathcal{N}^r_t(i)} w_{i,q}  X_{q,t-j}\} + \lambda]^2\\
 &= ||\mathbf{Y}-X\mathbf{\theta}||^2,
\end{align} 
where $\mathbf{Y}$ is the flattened time series that is to be fitted, and $X$ is the corresponding design matrix.
More specifically,
\begin{equation}\label{Y}
\mathbf{Y} = (X_{1,p+1}, \ldots, X_{1,n}, \ldots, X_{N,p+1}, \ldots, X_{N,n})^T
\end{equation}
and design matrix $X$ is given by
\begin{equation}\label{X}
    \begin{bmatrix}
    X_{1,p} & S_{p+1,1,1,1} & \ldots & S_{p+1,1,s_1,1} & \dots & X_{1,1} & S_{p+1,1,1,p} & \dots & S_{p+1,1,s_p,p} & 1\\
    \vdots &&&&&\vdots&&&&\vdots\\
    X_{N,p} & S_{p+1,N,1,1} & \ldots & S_{p+1,N,s_1,1} & \dots & X_{N,1} & S_{p+1,N,1,p} & \dots & S_{p+1,N,s_p,p} & 1\\
    \vdots &&&&&\vdots&&&&\vdots\\
    X_{1,n-1} & S_{n,1,1,1} & \ldots & S_{n,1,s_1,1} & \dots & X_{1,n-p} & S_{n,1,1,p} & \dots & S_{n,1,s_p,p} & 1\\
    \vdots &&&&&\vdots&&&&\vdots\\
    X_{N,n-1} & S_{n,N,1,1} & \ldots & S_{n,N,s_1,1} & \dots & X_{N,n-p} & S_{n,N,1,p} & \dots & S_{n,N,s_p,p} & 1\\\\
\end{bmatrix},
\end{equation}
where $S_{t,i,r,j} = \sum_{q\in \mathcal{N}^r_{t}(i)} w_{i,q}  X_{q,t-j}$.

In practice, we use the constrained least squares algorithm described by
\cite{BCL99} and implemented by the \verb+scipy.optimize.lsq_linear+ function from the {\tt Scipy}
python package, see \cite{2020SciPy-NMeth} with constraints of $[0, 1]$ on the individual $\alpha$ and
$\beta$ parameters. Let $\hat{\mathbf{\theta}}^{(n)}_{[0,1]} = \argmin_{\theta} Q_n(\theta)$ be the constrained estimator.


\subsubsection{Asymptotic properties}
\label{sec:gnariasym}

Let $\mathbf{X}_t = (X_{1, t}, X_{2, t}, \ldots, X_{N, t})^T$ is to be considered as a column vector with
components that are a stationary GNARI process as defined in~\eqref{eq:gnari}.
\begin{definition}
Define $\hat{\mathbf{X}}_{t|t-1}(\mathbf{\theta}) = \mathbb{E}_{\mathbf{\theta}}(\mathbf{X}_t|\mathcal{F}_{t-1})$
and
\begin{equation}\label{eq:f}
f_{t|t-1}(\mathbf{\theta}) = \mathbb{E}[ \{ \mathbf{X}_t -\hat{\mathbf{X}}_{t|t-1}(\mathbf{\theta}) \} \{\mathbf{X}_t-\hat{\mathbf{X}}_{t|t-1}(\mathbf{\theta})\}^T|\mathcal{F}_{t-1}]
\end{equation}
and let $\mathbf{\theta}_0$ be the true value of $\mathbf{\theta}$. 
\end{definition}
We now show asymptotic consistency and that the estimator is asymptotically normal.
\begin{proposition}
\label{prop:gnariasym}
    Assuming GNARI process stationarity, then

    \begin{enumerate}
        \item \begin{equation}
            \hat{\theta}^{(n)} \xrightarrow{a.s} \mathbf{\theta}_{0}.
        \end{equation}
        \item \begin{equation}
    n^{1/2}(\hat{\mathbf{\theta}}^{(n)}-\mathbf{\theta}_0) \xrightarrow{} \MVN(0,U^{-1}RU^{-1}),
\end{equation}
as $n\xrightarrow{} \infty$, where $\MVN$ is the multivariate normal distribution and
\begin{equation}
    U = \mathbb{E} \left\{ \frac{\partial \hat{\mathbf{X}}_{t|t-1}^T}{\partial \mathbf{\theta}}(\mathbf{\theta}_0)\frac{\partial \hat{\mathbf{X}}_{t|t-1}}{\partial \mathbf{\theta}}(\mathbf{\theta}_0) \right\},
\end{equation}
and
\begin{equation}\label{eq:R}
    R =  \mathbb{E}\left\{ \frac{\partial \hat{\mathbf{X}}_{t|t-1}^T}{\partial \mathbf{\theta}}(\mathbf{\theta}_0)f_{t|t-1}(\mathbf{\theta}_0)\frac{\partial \hat{\mathbf{X}}_{t|t-1}}{\partial \mathbf{\theta}}(\mathbf{\theta}_0)\right\}.
\end{equation}
    \end{enumerate}

\end{proposition}

See appendix for proof.

\begin{proposition}\label{prop:gnariasym0,1}
    Assuming GNARI process stationarity, then

    \begin{enumerate}
        \item \begin{equation}
            \hat{\theta}^{(n)}_{[0,1]} \xrightarrow{a.s} \mathbf{\theta}_{0}.
        \end{equation}
    \end{enumerate}
\end{proposition}

\subsubsection{Predictions}

Suppose we have a set of estimated parameters $\hat{\alpha}_{i}$, $\hat{\beta}_{i,j}$, and $\hat{\lambda}$, then, conditional on $\mathcal{F}_{n}$, the predicted mean of $X_{i,n+1}$ is simply given by 
\begin{equation}
\label{eq:GNARIpred}
    \hat{\mathbb{E}}(X_{i,n+1}|\mathcal{F}_n) = \sum_{j=1}^{p} \{\hat{\alpha}_{j}  X_{i, n-j+1} +\sum_{r=1}^{s_j} \hat{\beta}_{j,r}  \sum_{q\in \mathcal{N}^r_t(i)} (w_{i,q}  X_{q,n-j+1})\} + \hat{\lambda}.
\end{equation}
Further future predictions can be computed by recursing~\eqref{eq:GNARIpred}.

\subsection{The NGNAR model}

The NGNAR model adapts the GAR model from Section~\ref{sec:review} to
networks. The relationship
between NGNAR and GNAR is similar to that between the generalized
and ordinary linear
models.

\subsubsection{NGNAR model definition}

The \mbox{$D$-NGNAR($p,[s_1, \ldots, s_p]$)} process has the following structure:
\begin{equation}
\label{eq:ngnar}
    \begin{gathered}
    X_{i,t}|\mathcal{F}_{t-1} \sim D(M_{i,t}), \\
 M_{i,t} = g \left\{ \alpha_{i,0}+\sum_{j=1}^{p} (\alpha_{i,j} X_{i, t-j} +\sum_{r=1}^{s_j} \beta_{j,r} \sum_{q\in N^r_t(i)} w_{i,q}^{(t)} X_{q,t-j}) \right\},
    \end{gathered}
\end{equation}
where  $\mathcal{F}_{t}$ is the \mbox{$\sigma$-algebra} from Section~\ref{sec:evcls},
$D(m)$ is some exponential family distribution  with mean $m$
and $g:\mathbb{R} \xrightarrow{} \mathbb{R}$ is the response function.
All  other specifications are as for the GNAR($p,[s_1, \ldots, s_p]$) model. As for GNAR and GNARI models
the parameter $\alpha_{i,j}$ is permitted to be global (not depend on $i$),
and it is also possible to drop $\alpha_{i,0}$.
A key feature of the NGNAR model is its ability to adapt to negative autocorrelations.

NGNAR models that are restricted to only having stage one neighbours and lag one autoregression
are examples of the broad class
of nonlinear network autoregressions introduced by \cite{nna}. However, NGNAR models have the ability to
model associations using more general autoregressive lags, $p$, and $r$-stage neighbours, which have
proved important and effective for good network {\em time series} modelling. The more general models require modelling tools
to select model order, as regular ARIMA$(p, d, q)$ models do, such as AIC, BIC or network auto- and partial autocorrelations and visualizations of these such as Corbit plots, see \cite{corbit23}.

The choice of response function $g$ is important as it directly affects the relationship
between each node at each time-step. Some useful choices include:
\begin{itemize}
    \item The identity response: reducing the model to regular GNAR.
    
    \item The exponential response: $g(x) = \exp(x)$: in which case
    the model is similar to the log-linear Poisson autoregression model in~\cite{loglinear},
    but replacing the $X$ by $\log (X + 1)$, to prevent explosion as noted on page~564 of \cite{loglinear}.

    \item The $\relu$ function: $g(x) = r(x) = \max{(x,0)}$.
    
    \item The $\softplus$ function: $g(x) = s_c(x) = c^{-1} \log\{1+\exp(c x)\}$.
    As $c \rightarrow \infty$, the $\softplus$ function becomes  $\relu$.
\end{itemize}
Our exposition below uses the $\softplus$ response function with $c=1$, i.e., $s_c(x) = s_1(x)$. 
Estimation can be performed either by conditional least squares or conditional maximum likelihood  and
both are discussed below.

\subsubsection{Stationarity and ergodicity for NGNAR}
We prove stationarity conditions for NGNAR processes with the $\softplus$ response function next.
\begin{lemma}\label{lem:ngnar-sty}
A sufficient, but  not necessary, condition for static-network NGNAR($p,[s_1, \ldots,s_p]$) processes, with
$\softplus$ response, to be stationary is:
\begin{equation} \label{ngnar_stat}
    \sum_{j=1}^p(|\alpha_{i,j}|+\sum_{r=1}^{s_r} |\beta_{j,r}|) < 1, \ \ \  \forall i =1, \ldots, N.
\end{equation}
\end{lemma}
The NGNAR autocovariance function(s) will typically not have a closed form for many response
functions.
However, for a near-linear response function, such as $\softplus$,
the NGNAR autocovariance
will not be very different from that of the equivalent GNAR process.

\subsubsection{Existence of the moments of NGNAR}
\begin{lemma}\label{lem:ngnar-moment}
    Assuming that $D$ is the Poisson distribution and that $X_{i,t}|\mathcal{F}_{t-1}$ are mutually independent, then (\ref{ngnar_stat}) is also a sufficient, but  not necessary, condition for static-network NGNAR($p,[s_1, \ldots,s_p]$) processes to have $\mathbb{E}[\prod_{i=1}^N X_{i,t}^{m_i}] < \infty$ for all $t \geq 0$, $m_i \geq 0$.
\end{lemma}

\subsubsection{Remarks on conditional least squares estimation}
Again, for notational simplicity, we consider the global $\alpha$ case.
Let $\mathbf{\theta} = (\alpha_{1},\beta_{1,1},\dots,\beta_{1,s_1},\dots,\alpha_{p},\beta_{p,1},\dots,\beta_{p,s_p},\alpha_{0})^T$ be the parameter of interest, $\mathbf{y}$ be the target vector as defined in~\eqref{Y}, and $X$ be the design matrix as defined in~\eqref{X}.
Then, the conditional least squares estimator, $\hat{\mathbf{\theta}}$, is the one that minimizes
$|| Y-\mathbf{g}(X\mathbf{\theta}) ||^2$,
where $\mathbf{g}(\mathbf{x}) = \{ g(x_1), g(x_2), \ldots\}^T$.

As the process is no longer linear by design, we can not use the usual linear least squares estimation method.
Instead, we can use numerical methods such as gradient descent or the ADAM optimization in~\cite{adam}.
Choosing the solution to $X^TX\theta = X^TY$ as the initial value can speed up the optimization.

\begin{proposition}\label{prop:ngnarasym_ls}
    Under the assumption of Lemma~\ref{lem:ngnar-moment}, we have

 \begin{enumerate}
        \item \begin{equation}
            \hat{\theta}^{(n)} \xrightarrow{a.s} \mathbf{\theta}_{0}.
        \end{equation}
        \item \begin{equation}
    n^{1/2}(\hat{\mathbf{\theta}}^{(n)}-\mathbf{\theta}_0) \xrightarrow{} \MVN(0,U^{-1}RU^{-1}),
\end{equation}
as $n\xrightarrow{} \infty$, where $\MVN$ is the multivariate normal distribution, $U$ and $R$ are analogous to that in Proposition~\ref{prop:gnariasym}.
    \end{enumerate}
\end{proposition}

\subsubsection{Quasi-maximum likelihood estimation}
Unlike the GNARI model, the NGNAR model explicitly defines the conditional distribution of $X_{i,t}|\mathcal{F}_{t-1}$. Thus,
it is feasible to implement the quasi-maximum likelihood estimator (QMLE).
The  QMLE  $\hat{\mathbf{\theta}}_M$ maximizes the quasi-likelihood
\begin{equation}
    L(\mathbf{\theta}) =
    \prod_{t=p}^{n-1}\prod_{i=1}^N f(X_{i,t+1}|\mathcal{F}_{t},\mathbf{\theta}),
\end{equation}
where $f(\mathbf{X_{i,t+1}}|\mathcal{F}_{t},\mathbf{\theta})$ is the density function for $X_{i,t+1}|\mathcal{F}_t$. We also have that $\mathbb{E}[\mathbf{X_{i, t+1}}|\mathcal{F}_{t}, \mathbf{\theta}] = g([X]_{t-p+i} \cdot \mathbf{\theta})$, where $[X]_{r}$ is the $r^{th}$ row of matrix $X$ defined in \eqref{X}.
We can then recognize that the objective function under the NGNAR model assumption is of the same form as that of a generalized linear model (GLM). Thus, the solution for $\hat{\mathbf{\theta}}_M$ can be computed using similar techniques as used for GLM, such as iterative weighted least squares.

If we assumed that the conditional distribution was Poisson,
then the parameters estimated by conditional least squares would differ from those estimated by
QMLE unless $X_{i,t}$ is large enough so that the Poisson conditional distribution can be approximated with a Gaussian conditional distribution. Then the two estimation methods should give similar results. If $X_{i,t}|\mathcal{F}_{t-1}$ are mutually independent, the quasi-log-likelihood is

\begin{equation}
    l_n(\theta) = \sum_{t=p+1}^n \sum_{i=1}^N X_{i,t} \log(M_{i,t}) - M_{i,t} - \log(X_{i,t}!)
\end{equation}

\begin{proposition}\label{prop:ngnarasym_mle}
    Under the assumption of Lemma~\ref{lem:ngnar-moment}, we have

 \begin{enumerate}
        \item \begin{equation}
            \hat{\theta}^{(n)}_M \xrightarrow{a.s} \mathbf{\theta}_{0}.
        \end{equation}
        \item \begin{equation}
    n^{1/2}(\hat{\mathbf{\theta}}^{(n)}-\mathbf{\theta}_0) \xrightarrow{} \MVN(0,U^{-1}RU^{-1}),
\end{equation}

where $U$ and $R$ are defined in the proof in the appendix.
    \end{enumerate}
\end{proposition}

\subsubsection{Predictions}

Using either of the above
estimation methods
that obtain estimated parameters, $\hat{\alpha}_{i}$, $\hat{\beta}_{i,j}$, and $\hat{\alpha}_0$, the predicted mean of $X_{i,n+1}|\mathcal{F}_{n}$ is then
\begin{equation}
    \hat{\mathbb{E}}[X_{i,n+1}|\mathcal{F}_n] = g[\sum_{j=1}^{p} \{\hat{\alpha}_{j}  X_{i, n-j+1} +\sum_{r=1}^{s_j} \hat{\beta}_{j,r}  \sum_{q\in \mathcal{N}^r_t(i)} (w_{i,q}  X_{q,n-j+1})\} + \hat{\alpha_0}],
\end{equation}
which can clearly be computed
recursively for further horizons.

\section{Simulation Studies}

\subsection{GNARI parameter estimation simulation study}

\begin{figure}
    \centering
    \begin{minipage}[c]{0.45\textwidth}
        \includegraphics[width=\linewidth]{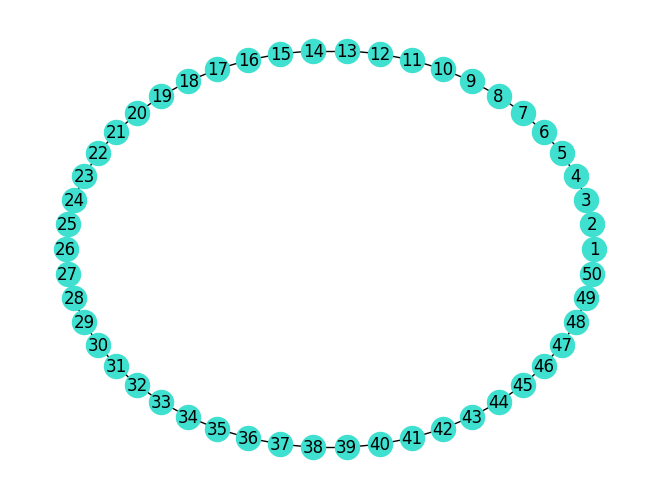}
   \end{minipage}
\hfill
    \begin{minipage}[c]{0.45\textwidth}
    \begin{equation*}
    M =\begin{bmatrix}
     0&1& &&&1\\
     1&0&1&& &\\
     &&\ddots&&&\\
     &&&\ddots&&\\
     &&&1&0&1\\
     1&&&&1&0
\end{bmatrix}
\end{equation*}
   \end{minipage}
           \caption{50-node chain network for simulation experiments. Left: picture of the network. Right: adjacency matrix.\label{fig:networkM}}
\end{figure}

This section investigates estimation performance using
conditional least squares for a
Poisson-GNARI($1,[1]$) process with $\alpha_1 = 0.5$, $\beta_{1,1} = 0.4$, and $\lambda =10$ on a $N=50$ chain network
shown with its adjacency matrix in
Figure~\ref{fig:networkM}.
We will simulate realizations from the Poisson-GNARI($1, [1]$)
process with lengths $T = 10, 50, 200$ and $500$
observations and repeat this $1000$ times
for each choice of $T$ and estimate
parameters for each realization.
\begin{table}
    \centering
    \begin{tabular}{c|c|c|c}
       $T$  &$\alpha_1$& $\beta_{1,1}$ & $\lambda$\\
         \hline
10 & 0.494 (0.039) & 0.392 (0.053) & 11.48 (5.08)\\
50 & 0.497 (0.017) & 0.397 (0.021) & 10.61 (1.99)\\
200& 0.500 (0.0080) & 0.399 (0.010) & 10.15 (0.93)\\
500& 0.500 (0.0053) &  0.400 (0.0070) & 10.07 (0.63)\\\hdashline
True & 0.500 & 0.400 & 10.0
    \end{tabular}
    \caption{The mean (and standard deviation) over $1000$ conditional least squares estimates of each parameter in the GNARI($1,[1]$) model for each length $T$}
    \label{tab:gnari_sim}
\end{table}
\mbox{Table~\ref{tab:gnari_sim}} shows the
results: the mean of the estimates
clearly approaches the truth as $T$
gets larger.
The standard deviation is approximately inversely proportional to $\sqrt{T}$, i,e., $\frac{\text{std}(\hat{\theta}_T)}{\text{std}(\hat{\theta}_m)} \approx \sqrt{\frac{m}{T}}$, which  is consistent with the asymptotic properties.
It is also worth noting that there is a tendency for underestimation
of $\alpha_1$ and $\beta_{1,1}$,
but overestimation of $\lambda$.

\subsection{NGNAR parameter estimation simulation study}
Table~\ref{tab:ngnar_sim} shows the results
of a similar simulation
study to the previous one, but for
a NGNAR process.
We simulate 1000 realizations
of a Poisson-NGNAR($1,[1]$) with $g(\cdot)= \softplus(\cdot)$, $\alpha_{1}=0.5$, $\beta_{1,1}=-0.4$, and $\alpha_0=10$ for 
lengths $T=10, 50, 200, 500$ on the
same network as in the previous section.

For each realization, we fit the NGNAR model
using both conditional least squares and conditional maximum likelihood. 
\begin{table}
    \centering
    \begin{tabular}{c|c|c|c|c}
       $T$ & Est.\ Method &$\alpha_1$& $\beta_{1,1}$ & $\alpha_0$\\
         \hline
10 &CLS & 0.494&-0.399&10.06  \\
& & (0.039)&(0.049)&(0.848)\\
 &CMLE & 0.494& -0.399 &10.06\\
&&(0.038) &(0.047)&(0.847)\\
\hline
50 &CLS & 0.499&-0.400&10.0  \\
& & (0.017)&(0.021)&(0.364)\\
 &CMLE & 0.499& -0.400 &10.0\\
&&(0.017) &(0.020)&(0.361)\\
\hline
200& CLS&0.500&  -0.400&  10.0\\
&&(0.0088)& (0.010)& (0.183)\\
 & CMLE& 0.500 &-0.400&  10.0\\
&&(0.0085)& (0.0098) &(0.181)\\
\hline
500& CLS&0.500& -0.400&  10.0\\
&&(0.0053) &(0.0066)& (0.115)\\
& CMLE&0.500& -0.400&  10.0\\
&&(0.0051) &(0.0062)& (0.112)\\\hdashline
True & & 0.500 & -0.400 & 10
\end{tabular}
\caption{The mean (and standard deviation) across the 1000 conditional least squares and conditional MLE estimates for each parameter in the Poisson-NGNAR($1,[1]$) model for each length $T$.}
    \label{tab:ngnar_sim}
\end{table}

\subsection{Predictive comparison via simulation}
We compare our new
GNARI and NGNAR models with the
recent PNAR
count network time series model
of~\cite{pnar}
via simulation.
Using the same network as earlier (as shown in Figure~\ref{fig:networkM})
we
simulate $500$ realizations of length $T=500$
from each of the following processes
\begin{description}
    
    \item [P1] Poisson-GNARI(1,[1]) with $\alpha_1 = 0.5$, $\beta_{1,1} = 0.4$, and $\alpha_0=10$;
    
    \item [P2] Poisson-NGNAR(1,[1]) with $g(\cdot) = \softplus(\cdot)$, $\alpha_1 = 0.5$, $\beta_{1,1} = 0.4$, $\alpha_0 = 10$;
    
    \item [P3] Poisson-NGNAR(1,[1]) with $g(\cdot) = \softplus(\cdot)$, $\alpha_1 = 0.1$, $\beta_{1,1} = -0.8$, $\alpha_0 = 10$;
    
    \item [P4] PNAR(1) with $\alpha_1 = 0.5$, $\beta_{1} = 0.4$, $\beta_0 = 10$.
\end{description}
For each simulated realisation,
we fit the following models: (A) GNARI($1,[1]$),
(B) NGNAR($1,[1]$) fitted by conditional least squares
and (C) by
conditional maximum likelihood, (D) PNAR($1,[1]$).

For this study we are interested in how well the models perform in terms of predictive performance.
To do this, we divide each network time series into a training set of length $450$ and a test set of length $50$. We fit (A) to (D) on the training set, and then make a prediction of length 50, which is then compared to the test values
and is assessed using mean-squared prediction error (MSPE).

\begin{table}
\centering
\begin{subtable}{.5\textwidth}
\centering
\begin{tabular}{c|c|c|c|c}
    & \multicolumn{4}{c}{Simulated Process}\\
     & P1 & P2& P3 & P4\\
     \hline
   A &67.2 & 99.9& 9.59& 99.6\\ 
   B & 67.2& 99.9 &5.89 &99.6\\ 
   C &67.2 &99.9 &5.89 & 99.6\\ 
   D &67.2 &99.9 & 9.59 &99.6\\ 
\end{tabular}
\caption{$h=1$}
\end{subtable}
\hfill
\begin{subtable}{.5\textwidth}
\centering
\begin{tabular}{c|c|c|c|c}
        & \multicolumn{4}{c}{Simulation Process}\\
     & P1 & P2& P3& P4\\
     \hline
   A &111.4& 166.3& 10.19  &166.6\\ 
   B & 111.4& 166.3 &9.06 &166.6\\ 
   C &111.4 &166.3 &9.06  &166.6\\ 
   D &111.4 &166.3 & 10.19 &166.6\\ 
\end{tabular}
\caption{$h=10$}
\end{subtable}
\hfill
\begin{subtable}{.5\textwidth}
\centering
\begin{tabular}{c|c|c|c|c}
        & \multicolumn{4}{c}{Simulated Process}\\
     & P1 & P2 & P3 & P4\\
     \hline
   A &129.8 & 192.6& 10.31&194.5\\ 
   B & 129.8& 192.6 &10.07 &194.5\\ 
   C &129.8 &192.6 &10.07 &194.5\\ 
   D &129.8 &192.6 & 10.31&194.5\\ 
\end{tabular}
\caption{$h=50$}
\end{subtable}

\caption{Average mean-squared prediction error (MSPE)
between the predicted next $h$ days time series by each model (A, B, C, D) fitted to
training realizations from simulated processes P1, P2, P3 and P4 on the training set evaluated on the test set
over 500 realizations from each simulated model.
\label{tab:comparison}}
\end{table}

Table \ref{tab:comparison} shows that for the simulated GNARI, NGNAR with positive parameters and PNAR processes,
all four methods have almost equal performance.
For the NGNAR simulation with $\beta_{1,1} = -0.8$, the NGNAR models definitively predict better than the other models, especially
over the short term. However, for longer horizons there is not much to choose between the methods.

The NGNAR model is clearly
more flexible than the GNARI model and the PNAR model as it can cope with negative parameters.
In principle, the log-linear version of
the \cite{pnar} model can cope with negative
parameters. However, the error
structure that the log-linear
model assumes is 
different to that of
the simulated processes
P1--P4 above. To avoid doubt
we repeated the simulation/prediction
exercise above for the log-linear
model and the prediction errors
were uniformly at least four times
worse than those reported
in Table~\ref{tab:comparison} and,
in many cases, much worse.


\section{Example: New York State COVID forecasting}

\subsection{Data description}
We obtained daily counts of people
who tested positive for COVID19 in
the 62 counties of New York State, USA
from~\cite{NYcovid} during
the period 1st March 2020
to 23rd May 2022. 
The counts can be written as a multivariate time
series of dimension $T\times N=783\times 62$
(some common missing values at the head of the time series
were discarded).

\begin{figure}
    \centering
    \resizebox{0.65\textwidth}{!}{\includegraphics{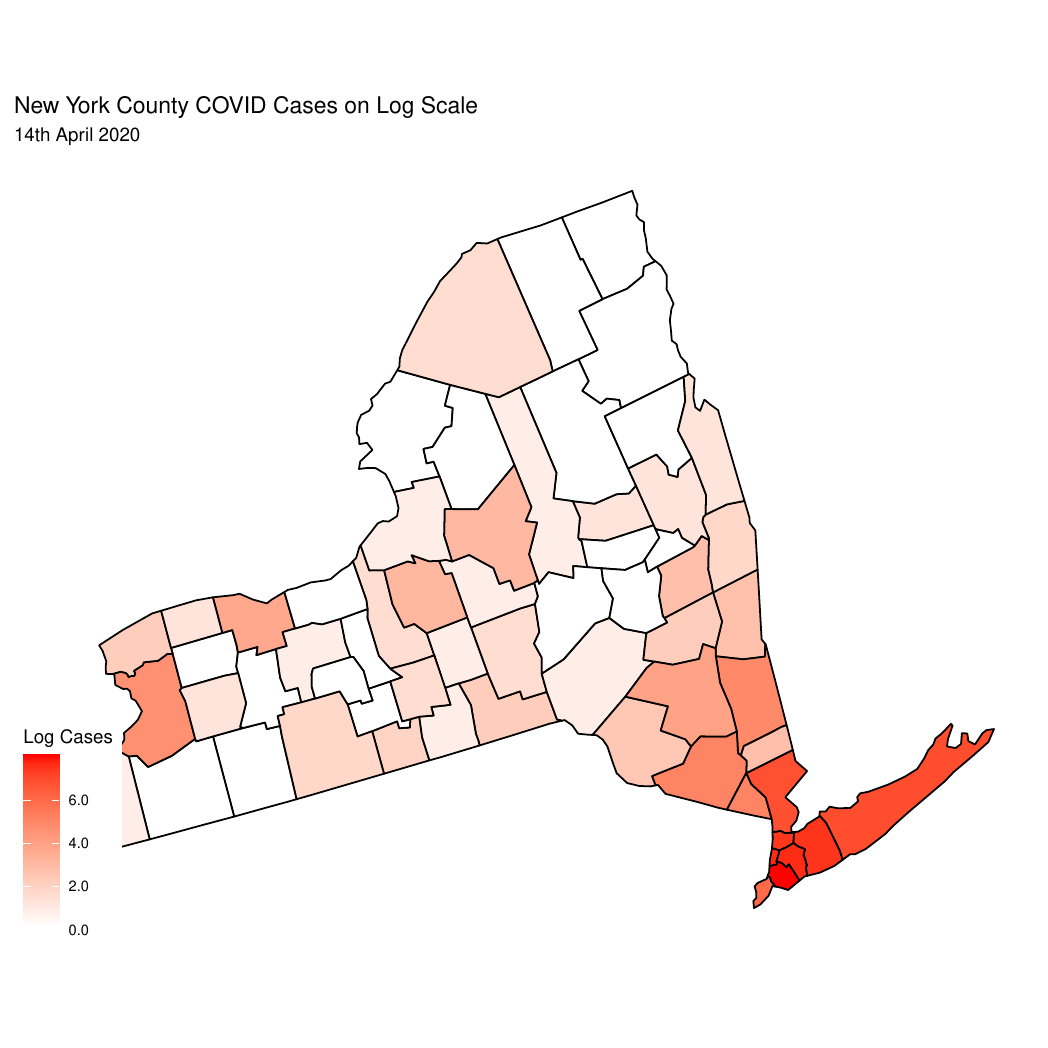}}
    \caption{Counties of New York
        State with log COVID cases indicated by heat map.
    \label{fig:nystate}}
\end{figure}

\begin{figure}
    \centering
    \resizebox{0.65\textwidth}{!}{\includegraphics{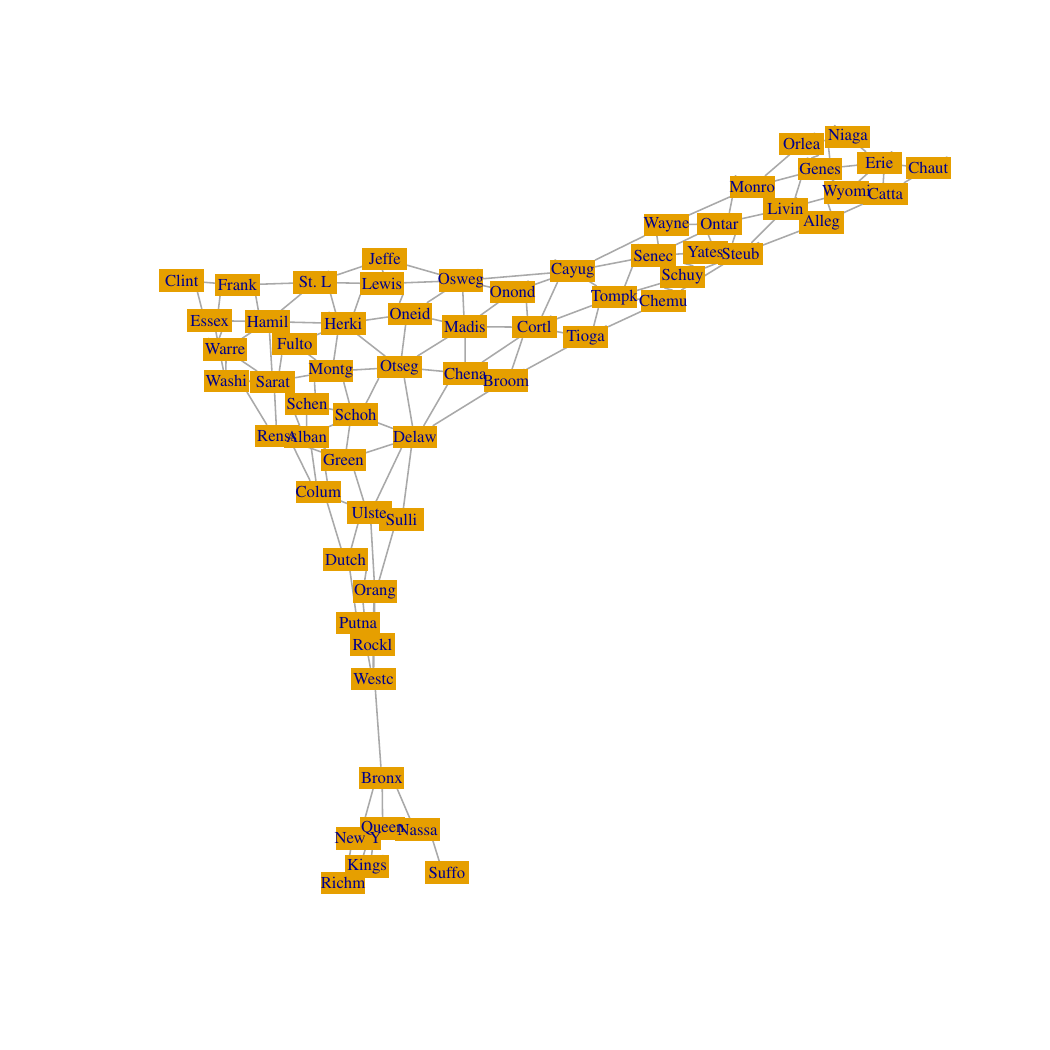}}
    \caption{Graph associated with
        New York state counties.
    \label{fig:nynet}}
\end{figure}

Figure~\ref{fig:nystate} shows
a map of the counties of New York
State, which are colour-coded
according to the logarithm of the
number of COVID positives. It
can be seen that the highest
count is centred in and around
New York City at the bottom right
of the map.
We constructed a network for
the counties by treating each
county as a vertex and joining
two vertices if the respective
counties shared a border.
Network weights are equally allocated
between the neighbours
(e.g.\ if a county has $k$ neighbours,
then the out-weight of that
county to each of its neighbours
is $k^{-1})$.
Figure~\ref{fig:nynet} depicts
the graph we use for our
network time series modelling.
The New York city area can
be seen at the `bottom' of the graph
in
Figure~\ref{fig:nynet}.

\subsection{Prediction evaluation method}
We split the multivariate time series into a training series of length $700$ and a test series
of length $83$. The time series for each county follows a similar pattern, but the actual
$X_{i, t}$ values are different in level. This indicates that we can use a global $\alpha$ 
in the model, but local values of $\lambda$ or $\alpha_0$ for the GNARI and NGNAR models,
respectively. In other words, the $\alpha_{i, 0}$ in~\eqref{eq:ngnar} and the
$\lambda_i$ as the expectation of the noise in~\eqref{eq:gnari} will both not be 
constant values of $i$.
The autocorrelations of $\{X_{i,t}\}_t$ and the cross-correlations between $\{X_{i,t}\}_t$ and its weighted sum $\{\mathcal{N}^{(1)}(X_{i,t-\tau})\}_t$, $\tau=0, 1, \ldots, 30$ are all positive:
at least for the first $30$ lags,
which shows the feasibility of the GNARI model.

We will fit four model types on the training series: GNAR, PNAR, GNARI and 
NGNAR models. Each model will be fitted twice using maximum lags of $14$ and $21$, respectively.
The order of $\alpha$, $[I_1,\dots,I_p]$ where $I_j$ is either 1 or 0 indicating whether the autoregressive term at lag $j$ is included. The order of $\beta$, $[s_1, \ldots, s_p]$, will be selected using backward deletion and the Bayesian information criterion (BIC) as the metric.
For the PNAR model, the quasi-MLE fitting method provided in the {\tt PNAR}
package in R can only fit a PNAR(1) model for this particular dataset as
higher order models will lead to a non-zero score function.

We will fit the GNAR model using conditional least squares,
the GNARI model using  constrained least squares to ensure that the model parameters are non-negative.
The NGNAR models will be fitted using conditional MLE using the ADAM optimizer from~\cite{adam}.
In the GNARI model we will assume that the $\epsilon_{i,t}$ is Poisson-distributed. The NGNAR models will use the $\softplus$ function as the response function and Poisson as the conditional distribution.

For each model, we predict the time series for the next 83 days. Denote the predicted time series $\{\hat{X}_{i,t}^{(M)}\}_{t=701}^{783}$ where $M$ is the model. The results will be shown by plotting $\{\hat{X}_{i,t}^{(M)}\}_{t=701}^{783}$ along with the true actual values $\{ X_{i,t}\}_{t=701}^{783}$ for some $i$. We will compute the mean-squared prediction error (MSPE) and mean absolute prediction error (MAPE)
between the next $T$ days prediction for each model and its corresponding true value, i.e., the MSPE between $\{\hat{X}_{i,t}^{(M)}\}_{t=701}^{700+h}$ and $\{X_{i,t}\}_{t=701}^{700+h}$, for $h=1, \ldots, 83$.

\subsection{Results}
\begin{figure}
    \centering
    \resizebox{\textwidth}{!}{\rotatebox{270}{\includegraphics{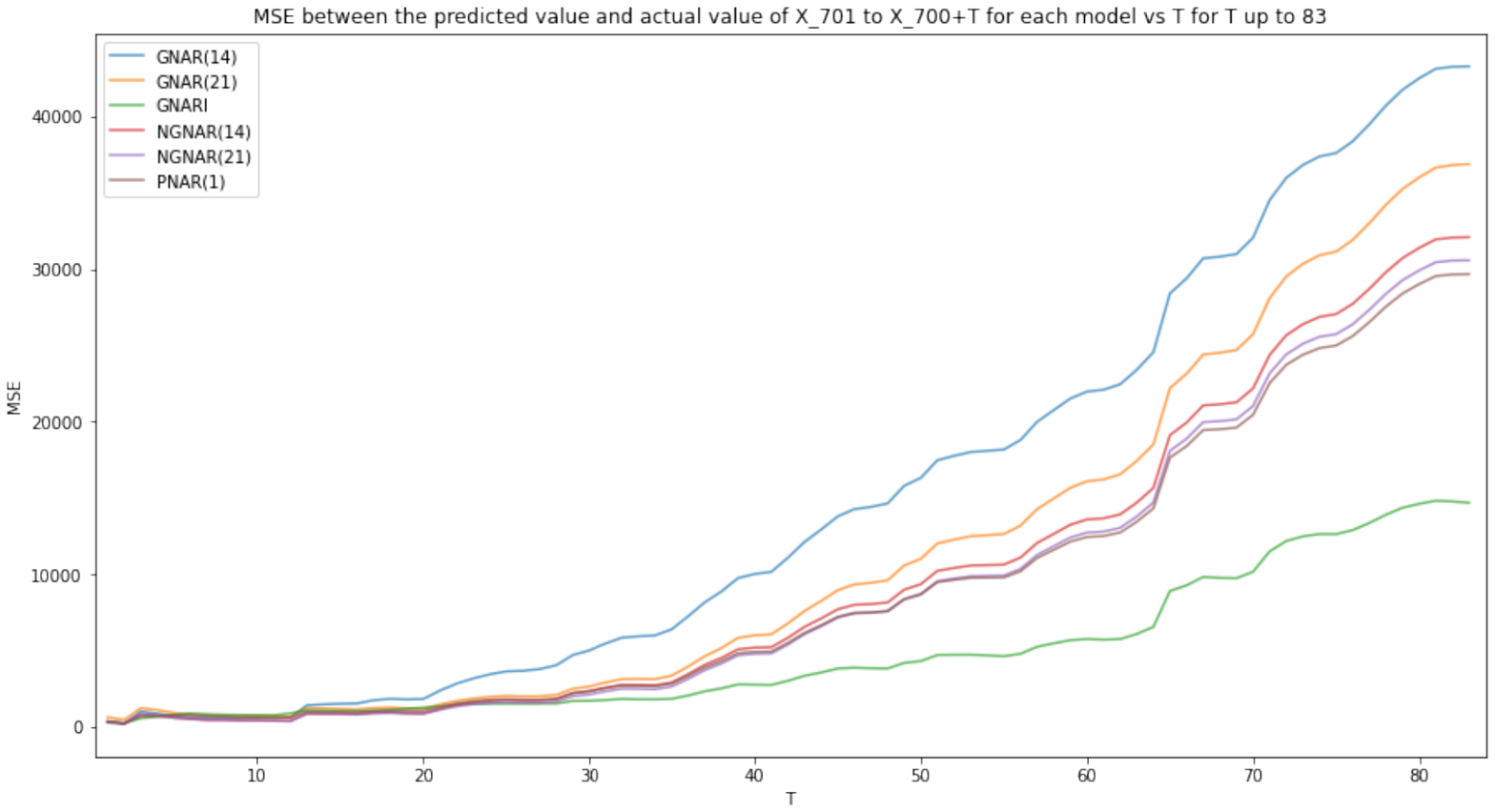}}}
    \caption{Plot of MAPE between the prediction of the next $h$ days time series and the corresponding true values, i.e., the aggregate MAPE between $\{\hat{X}_{i,t}^{(M)}\}_{t=701}^{700+h}$ and $\{X_{i,t}\}_{t=701}^{700+h}$, against $T$ for $T = 1, \ldots, 83$.}
    \label{fig:covidmape}
\end{figure}

The order selected for the GNARI($14$) and GNARI($21$) turned out to be identical, so only one
GNARI result is reported here.
Tables~\ref{tab:covidmspe},
\ref{tab:covidmape} and
Figure~\ref{fig:covidmape} clearly
show the superiority of the NGNAR
models (particularly the one of order
$14$) for short- to medium-term
forecasting and the GNARI model
for longer terms forecasts.
\begin{table}
    \centering
    \begin{tabular}{|r|r|r|r|r|r|r|}\hline
    & \multicolumn{6}{|c|}{Forecast horizon $h$}\\
    Model & $1$ & $5$ & $10$ & $25$ & $50$ & $83$ \\\hline
    GNAR($14$) & {\bf 3.18} & 7.47 & 6.55 & 36.6  & 163 & 433\\
    GNAR($21$) & 6.60 & 9.44 & 7.47 & 20.6 & 110 & 369\\
    GNARI      & 3.25 & 8.20 & 7.98 & {\bf 15.5} & {\bf 43.4} & {\bf 147}\\
    NGNAR($14$)& 3.85 & {\bf 6.25} & {\bf 4.53} & 17.9 & 93.7 & 321\\
    NGNAR($21$)& 3.97 & 6.33 & {\bf 4.53} & 16.5 & 87.3 & 306\\
    PNAR($1$)  & 3.95 & 7.91 & 6.44 & 18.4 & 87.0 & 297\\\hline
    \end{tabular}
    \caption{The mean-squared prediction error $\times 100$ (MSPE) between the predictions
        up to forecast horizon $h$ (to three significant figures). \label{tab:covidmspe}}
\end{table}

\begin{table}
\centering
    \begin{tabular}{|r|r|r|r|r|r|r|}\hline
    & \multicolumn{6}{|c|}{Forecast horizon $h$}\\
        Model & $1$ & $5$ & $10$ & $25$ & $50$ & $83$ \\\hline
    GNAR($14$) & 10.1 & 10.6 & 11.0 & 22.8 & 52.3 & 86.8\\
    GNAR($21$) & 12.9 & 12.4 & 12.2 & 17.3 & 39.6 & 74.5\\
    GNARI     & 10.4 & 13.9 & 14.1 & 18.3 & {\bf 26.4} & {\bf 45.6}\\
    NGNAR($14$)& {\bf 10.0} & {\bf 9.66} & {\bf 9.23} & 14.9 & 35.1 & 66.5\\
    NGNAR($21$)& 10.3 & 9.71 & 9.42 & {\bf 14.6} & 33.7 & 64.2\\
    PNAR($1$)  & 11.4 & 14.9 & 15.2 & 22.5 & 40.7 & 69.0\\\hline
    \end{tabular}
    \caption{The mean absolute prediction error (MAPE) between the predictions
        up to forecast horizon $h$ (to three significant figures). \label{tab:covidmape}}
\end{table}

\section{Discussion}
This article has introduced
two new models for network time
series that have count data as
observations. In general, the models are useful and work
well achieving similar performance to PNAR models for
data with positive autocorrelations. For the COVID data above
the new models performed particularly well and better than PNAR
and GNAR.
The NGNAR model
works well in estimating
negative network parameters,
unlike comparator models.
We have established the asymptotic
properties for GNARI and
NGNAR processes. For the latter 
showing its asymptotic normality by
utilising established results in this
context. Moreover, we have
described methods for estimation
using conditional least squares
and conditional maximum likelihood.

\section{Acknowledgements}
Liu gratefully acknowledges partial
support by Imperial College London.
Both gratefully
acknowledge support from
EPSRC NeST Programme grant EP/X002195/1.

\appendix 

\section{Proofs}
\begin{proof}[Proof of Lemma~\ref{lem:Z}]

We drop the $t$ subscript for simplicity.
The moment generating functions of $B_{j, r, k}$ and
$Y_{i, j}^{(r)}$ are
\begin{equation}
M_{B_{j, r, k}} (s) = 1 - \beta_{j, r} + \beta_{j, r} e^s
\end{equation}
and
\begin{equation}
M_{Y_{i, j}^{(r)}}(s) = \prod_{q \in \Ne_t^{(r)} (i)} (
    1-w_{i, q} - w_{i, q}e^s)^{X_{q, t-j}},
\end{equation}
for $s \in \reals$. For the moment generating function of $Z_{i, j}^{(r)}$ we have
\begin{align}
    M_{Z_{i, j}^{(r)}} (u) &=
    \E \left\{ e^{u Z_{i, j}^{(r)}} \right\}\\
    &= \E \left[ \E \{ e^{u Z_{i, j}^{(r)}} |
        Y_{i, j}^{(r)} \} \right]\\
    &= \E \{ (1 - \beta_{j, r} + \beta_{j, r} e^u )^{Y^{(r)}_{i, j}} \}\\
    &= M_{Y^{(r)}_{i, j}} \{ \log( 1-\beta_{j, r}
        + \beta_{j, r}e^u) \}\\
    &= \prod_{q \in \Ne_t^{(r)} (i)} \{1 - w_{i,q}
        + w_{i, q} (1 - \beta_{j, r} + \beta_{j, r}e^u)
        \}^{X_{q, t-j}}\\
    &=  \prod_{q \in \Ne_t^{(r)}} (1 -
        \beta_{j, r} w_{i, q} +
        \beta_{j, r} w_{i, q} e^u)^{X_{q, t-j}},
\end{align}
which is the moment generating function of the Poisson binomial distribution, with the parameters
specified in the statement of the lemma. By the uniqueness of MGFs, this is the distribution of the $Z^{(r)}_{i, j}$.

\end{proof}

\begin{proof}[Proof of Lemma~\ref{lem:sty}]

We prove the result for GNARI variant~\eqref{eq:gnari_variant}.
For $A \in \mathbb{R}^{N \times N}, \mathbf{X} = (X_1, \ldots, X_N) \in \mathbb{R}^N, N\in \mathbb{N}$,
define $A \circ \mathbf{X}$ to be 
\begin{equation}
    A \circ \mathbf{X} =\begin{pmatrix}
\sum_{k = 1}^{N} [A]_{1,k} \circ X_j \\
\vdots\\
\sum_{k = 1}^{N} [A]_{N,k} \circ X_j
\end{pmatrix}
\end{equation}
Hence, the GNARI process~\eqref{eq:gnari_variant} can be viewed as a multivariate integer-valued autoregressive
(MGINAR) process, see~\cite{mginar}:
\begin{equation}\label{mginarp}
    \mathbf{X}_t = \sum_{j=1}^p A_j \circ \mathbf{X}_{t-j} + \mathbf{\epsilon}_t,
\end{equation}
where $\mathbf{X}_t = (X_{1,t},\dots,X_{N,t})^T$, $A_j = \diag \{ \alpha_{i,j}\} + \sum_{r=1}^{s_j} \beta_{j,r}W^{(r)}$ and $W^{(r)}$ is the matrix with entries $$[W^{(r)}]_{l,m} = w_{l,m} \mathbb{I}\{ m \in \mathcal{N}^{(r)}(l)\}$$
and $\mathbf{\epsilon}_t = (\epsilon_{1,t},\ldots,\epsilon_{N,t})^T$.\\

Let 
\begin{equation}\label{mat_A}
    A = \begin{bmatrix}
A_1 & A_2 & \dots &A_{p-1}&A_p\\
I & 0 & \dots &0 &0\\
\vdots &\vdots &\ddots&\vdots&\vdots\\
0&0&\dots&I&0
\end{bmatrix}.
\end{equation}
Then, the GNARI process is equivalent to 
\begin{equation}\label{eq:mginar1}
    \mathbf{Y}_t = A \circ \mathbf{Y}_{t-1} + \mathbf{e}_t,
\end{equation}
where $\mathbf{Y}_t = (\mathbf{X}_t^T, \mathbf{X}_{t-1}^T, \ldots,\mathbf{X}_{t-p+1}^T)^T$
and $\mathbf{e}_t = (\epsilon_t^T, 0^T, \ldots, 0^T)^T$.\\

\cite{mginar} Proposition 3.1 permits us to conclude that, under the condition that all roots of $\det(I-Az)$
are outside the unit circle, or, equivalently, all eigenvalues of $A$ are inside the unit circle,
the process $\{\mathbf{X}_t\}_t$ has a unique stationary solution. \cite{mginar} Section 5 also shows that $\{\mathbf{X}_t\}_t$ is ergodic.

\cite{gnar} Appendix A tells us that the above condition can be achieved if condition~\eqref{eq:gnari_stat} holds. Hence, condition~\eqref{eq:gnari_stat} guarantees the unique stationary solution of the GNARI process.
\end{proof}

\begin{proof}[Proof of Proposition~\ref{prop:gnariasym}]
\label{proof:gnariasym}
    Assume that GNARI process is stationary. As in Section~\ref{sec:gnariasym} define 
\begin{equation}
    \hat{\mathbf{X}}_{t|t-1}(\mathbf{\theta}) = \mathbb{E}_{\mathbf{\theta}}[\mathbf{X_t}|\mathcal{F}_{t-1}].
\end{equation}
and let $\mathbf{\theta}_{0}$ be the true value of the (vector) parameter of interest~$\mathbf{\theta}$.
In the GNARI case, we have that 
\begin{equation}
    \hat{\mathbf{X}}_{t|t-1}(\mathbf{\theta}) = \sum_{j=1}^p A_j \mathbf{X_{t-j}} + \lambda\mathbf{1}_N.
\end{equation}
Then
\begin{equation}
    \label{eq:etalpha}
    \mathbb{E} \left[\left\|\frac{\partial\hat{\mathbf{X}}_{t|t-1}(\mathbf{\theta})}{\partial\alpha_{j}}\right\|^2\right] = \mathbb{E} [\mathbf{X}_{t-j}^T \mathbf{X}_{t-j}] := \eta_\alpha
\end{equation}
and
\begin{equation}
    \label{eq:etabeta}
    \mathbb{E}\left[\left\|\frac{\partial\hat{\mathbf{X}}_{t|t-1}(\mathbf{\theta})}{\partial\beta_{j,r}}\right\|^2\right] = \mathbb{E} [\mathbf{X}_{t-j}^T {W^{(r)}}^T W^{(r)}\mathbf{X}_{t-j}]
    := \eta_\beta,
\end{equation}
say, are expectations of quadratic forms in $\mathbf{X}_{t-j}$. By~\eqref{eq:mu_Y} and~\eqref{eq:Gamma_0}, we have both
$\eta_\alpha, \eta_\beta <\infty$.
Further, $\mathbb{E} \left[\left\| \partial\hat{\mathbf{X}}_{t|t-1}(\mathbf{\theta})/\partial\lambda\right\|^2\right] =  \mathbf{1}^T_N\mathbf{1}_N = N < \infty$.
We also know that all the
second-order derivatives with respect to $\alpha_j$, $\beta_{j,r}$, or $\lambda$  are all zero,
as $\hat{\mathbf{X}}_{t|t-1}$ depends linearly on each parameter.

Let $d$ be the length of the $\theta$ vector.
Suppose there exists constants $a_1, \dots, a_d$ such that
\begin{equation}
    \mathbb{E}\left[\left\| \sum_{k=1}^d a_k \frac{\partial\hat{\mathbf{X}}_{t|t-1}(\mathbf{\theta})}{\partial\theta_k} \right\|^2\right] = 0
\end{equation}
which implies that, for all $i=1, \dots, N$, there is a linear relationship between the $X_{q, t-j}$ given by 
\begin{equation}
    \sum_{j=1}^{p} \sum_{q \in \mathcal{N}_j(i)} C_{i,j,q} X_{q,t-j} + a_d = 0, \quad \quad a.s.
\end{equation}
where $\mathcal{N}_j(i) = {i}\cup\mathcal{N}^{(1)}(i)\cup \dots \cup \mathcal{N}^{(s_j)}(i)$, $C_{i,j,q} = c_{i,j,q}a_k$ for some $k$ and $c_{i,j,q}$ independent of $a_1,\dots,a_d$. This implies that $C_{i,j,q} = 0$, i.e., $a_1=\dots=a_d=0$.
Then, by Therorem~3.1 in~\cite{estnonlinear}, we know that $\hat{\theta}^{(n)} \xrightarrow{a.s} \mathbf{\theta}_{0}$.

Now, let 
\begin{equation}
    R =  \mathbb{E}\left[\frac{\partial \hat{\mathbf{X}}_{t|t-1}^T}{\partial \mathbf{\theta}}(\mathbf{\theta}_0)f_{t|t-1}(\mathbf{\theta}_{0})\frac{\partial \hat{\mathbf{X}}_{t|t-1}}{\partial \mathbf{\theta}}(\mathbf{\theta}_{0})\right],
\end{equation}
where as we assumed conditional independence,
\begin{align}
    f_{t|t-1}(\mathbf{\theta}) &= \mathbb{E}[\{ \mathbf{X_t}-\hat{\mathbf{X}}_{t|t-1}(\mathbf{\theta}) \} \{\mathbf{X_t}-\hat{\mathbf{X}}_{t|t-1}(\mathbf{\theta})\}^T|\mathcal{F}_{t-1}]\\
    &= \text{diag}(\{\textbf{Var}(X_{i,t}|\mathcal{F}_{t-1})\}_i),
\end{align}
with $\textbf{Var}(X_{i,t}|\mathcal{F}_{t-1})$ as in \eqref{gnari_cond_var}. In order to prove asymptotic normality, we require that $\|R\|_2 < \infty$. In our case, to temporarily simplify notation,
let $\theta_{a} = \beta_{j_1,r_1}$, $\theta_b = \beta_{j_2,r_2}$, then 
\begin{equation}
    R_{a,b} = \mathbb{E}\left[\sum_{i=1}^N \{ W^{(r_1)}X_{t-j_1}\}_{i} \textbf{Var}(X_{i,t}|\mathcal{F}_{t-1}) \{ W^{(r_2)}X_{t-j_2}\}_{i}\right]
\end{equation}
and similarly for the other entries. Since we have assumed that the $\epsilon_{i,t}$ are Poisson distributed,
we have $\mathbb{E}[|\epsilon_{i,t}|^3] < \infty$, which implies that $\mathbb{E}[|X_{i,t}|^3] < \infty$, see~\cite{franke1993multivariate}. We can then use the Cauchy-Schwarz inequality to show that $|R_{a,b}| < \infty$,
for all possible pairs $a, b$.

Hence, \cite{estnonlinear} Theorem~3.2 implies that as $n \xrightarrow{} \infty$ we have
\begin{equation}
    n^{1/2}\{ \hat{\mathbf{\theta}}^{(n)}-\mathbf{\theta}_{0} \} \xrightarrow{} \MVN(0, U^{-1}RU^{-1}),
\end{equation}
where 
\begin{equation}
    U = \mathbb{E}\left[\frac{\partial \hat{\mathbf{X}}_{t|t-1}^T}{\partial \mathbf{\theta}}(\mathbf{\theta}_0)\frac{\partial \hat{\mathbf{X}}_{t|t-1}}{\partial \mathbf{\theta}}(\mathbf{\theta}_0)\right]. 
\end{equation}
\end{proof}

\begin{proof}[Proof of Proposition~\ref{prop:gnariasym0,1}]
    The proof of \cite{estnonlinear} Theorem~3.1 shows that the regularity conditions of Theorem~3.1 which we have checked in our case as above implies the conditions of Theorem~2.1. The conclusion that $\hat{\theta}^{(n)} \xrightarrow{a.s} \mathbf{\theta}_{0}$ is a direct result of Theorem~2.1. We now show that Theorem~2.1 holds if we replace $\hat{\theta}^{(n)}$ with $\hat{\theta}^{(n)}_{[0,1]}$.
    
    First note that $Q_n(\theta)$ is globally convex in our case, so we do not need to consider local minima. The proof of \cite{estnonlinear} Theorem~2.1 states that for any $\epsilon, \delta >0$, there exists an event $E$ with $P(E) > 1-\epsilon$ and $n_0 \in \mathbb{N}$ such that on $E$, for any $n > n_0$ and $\theta$ on the boundary of $B_{\delta^*}(\theta^0)$ where $0 < \delta^* < \delta$ and $B_{\delta^*}(\theta^0)$ is the open sphere of radius $\delta^*$ centered at $\theta^0$,
    \begin{equation}
        Q_n(\theta) \geq Q_n(\theta^0)
    \end{equation}

    Moreover, the minimum $\hat{\theta}^{(n)}$ is in $B_{\delta^*}(\theta^0)$. This implies that $\hat{\theta}^{(n)}_{[0,1]}$ is in $B_{\delta^*}(\theta^0) \cap [0,1]^d$ on event $E$. The rest is identical to the proof of Corollary~2.1 \cite{klimko1978}.
\end{proof}

\begin{proof}[Proof of Lemma~\ref{lem:ngnar-sty}]
For the convenience of notation, we assume that $\alpha_{i,0} = 0$ for all $i$.\\

Let
\begin{equation}
\mathbf{Y}_t = (X_{1, t}, \ldots, X_{N, t}, \ldots, X_{1, t-p+1}, \ldots, X_{N, t-p+1})^T,
\end{equation}
where $A$ is the same matrix as in~\eqref{mat_A}, and $g(x) = s(x)$ is the entry-wise $\softplus$ function.
Then $\{\mathbf{Y}_t \}_t$ is Markov, aperiodic and irreducible and we have
\begin{equation}\label{eq:ngnar_markov}
    \mathbb{E}(\mathbf{Y}_t| \mathbf{Y}_{t-1}) = f(A\mathbf{Y_{t-1}}),
\end{equation}
where $f(\mathbf{X}) = \{ \underbrace{g(\mathbf{X}_1)^T}_{N}, \ldots, g(\mathbf{X}_p)^T, \mathbf{X}_{p+1}^T, \ldots\}^T $.
From Appendix~A in~\cite{gnar}, we know that (\ref{ngnar_stat}) implies that all eigenvalues of $A$ are
inside the unit circle. This is equivalent to the spectral radius of $A$, $\rho(A) < 1$. Thus, by Lemma~2.5 in~\cite{ergodicitynlar}, there exists a matrix norm $||\cdot||_m$, a vector norm $||\cdot||_v$, and $\lambda \in (0,1)$, such that 
\begin{equation}
    ||Ax||_v \leq ||A||_m||x||_v \leq \
\lambda ||x||_v, \ \ \ \forall x \in \mathbb{R}^{Np}.
\end{equation}
Now, $\forall \mathbf{y} \in \mathbb{R}^{Np}$

\begin{equation}
    \mathbb{E}( ||\mathbf{Y}_{t+1}||_v \, |\, \mathbf{Y}_t=\mathbf{y}) \leq  ||f(A\mathbf{y})||_v
\end{equation} 
and
\begin{equation}
    ||A\mathbf{y}||_v \leq \
\lambda ||\mathbf{y}||_v.
\end{equation}
It is simple to show that 
 $\lim_{x \xrightarrow{} \infty}g(x) = x$, and $|g(x)| < |x|$ for all $x < \arcsinh(-1/2)$, for $g(x) = s_1(x)$, the
 $\softplus$ function. We can thus find $C \in \mathbb{R}^{Np}$ to be $R^{Np}_{\geq 0} \bigcap C'$ where $C'$ is a sphere in $\mathbb{R}^{Np}$ such that if $\mathbf{y} \notin C$, the negative elements of $A\mathbf{y}$ are less than $\text{archsinh}(-1/2)$ and $||\mathbf{y}'-A\mathbf{y}||_v < (1-\lambda) M/2$, where $\mathbf{y}'$ is the vector whose non-negative entries equals that of $f(A\mathbf{y})$ and negative entries equals that of $A\mathbf{y}$, and $M=\inf_{\mathbf{y} \notin C}||Y||_v$. \\

Then $\forall \mathbf{y} \notin C$, 
\begin{align*}
    ||f(A\mathbf{y})||_v &\leq ||A\mathbf{y}'||_v \\
    &\leq ||\mathbf{y}'-A\mathbf{y}||_v + ||A\mathbf{y}||_v\\
    &\leq (1-\lambda) M/2 + \lambda ||\mathbf{y}||_v\\
    &\leq (1/2+\lambda/2)||\mathbf{y}||_v
\end{align*}
where clearly $1/2+\lambda/2<1$.\\

Now, by Lemma~2.2 from~\cite{ergodicitynlar}, which is a reformulation of the Tweedie's criterion for ergodicity (\cite{Tweedie1975SufficientCF}):\\

\emph{
Let $\{\mathbf{Y_t}\}$ be aperiodic irreducible. Suppose
that there exist a small set $C$, a nonnegative measurable function $g$, positive constants $c_1$, $c_2$ and $\rho < 1$ such that}

\begin{equation}
    \mathbb{E}[g(\mathbf{Y_{t+1}})|\mathbf{Y_t}=\mathbf{y}]\leq \rho g(\mathbf{y})-c_1, \ for \ any \ y\notin C
\end{equation}

\emph{and} 
\begin{equation}
    \mathbb{E}[g(\mathbf{Y_{t+1}})|\mathbf{Y_t}=\mathbf{y}]\leq c_2, \ for \ any \ y\in C
\end{equation}
\emph{Then $\{\mathbf{Y_t}\}$ is geometrically ergodic.}\\
Also, the fact that the $\softplus$ function is bounded for the bounded region, we have that the Markov process $\{\mathbf{Y_t}\}_t$ is geometrically ergodic and has a unique stationary solution. This implies that the NGNAR process $\{X_t\}_t$ has a unique stationary solution.
\end{proof}

\begin{proof}[Proof of Lemma~\ref{lem:ngnar-moment}]

Let $v^{a*}$ denote the entry-wise $a^\text{th}$ exponential of a vector $v$. Let $*$ be the entry-wise multiplication. Let $\leq^*$ be the entry-wise comparison. Let $A$ be any matrix, define $|A|$ to be the matrix with $|A|_{i,j} = |A_{i,j}|$, similarly for vectors. For notation convenience, we prove for the case where $p=1$.

First note that $s(x) \leq \log(2) + x^+$ for all $x$ where $x^+ = x \mathbf{I}_{x\geq 0}$. We also know that the $m^\text{th}$ moment of a Poisson$(\lambda)$ random variable is $\sum_{u=0}^m \bracenom{m}{u} \lambda^u$ where $\bracenom{m}{u}$ is the Sterling number of the second kind. Let $A_1$ be as in (\ref{mginarp}). Let $\alpha_0 = (\alpha_{1,0},\dots,\alpha_{N,0})^T$. For any $k \geq 0$, we thus have that

\begin{align}\label{eq:ngnar-cond_moment}
    \mathbb{E}[\mathbf{X}_t^{k*}|\mathcal{F}_{t-1}] &= \sum_{u=0}^k \bracenom{k}{u} (s(A_1\mathbf{X}_{t-1}))^{k*}\\
    &\leq^*  \sum_{u=0}^k \bracenom{k}{u} \sum_{l=0}^u (\log(2) + |\alpha_{0}|)^{(u-l)*} *(|A_1|\mathbf{X}_{t-1})^{l*}
\end{align}

We have also assumed that\label{eq:ngnar-cond_indep}

\begin{equation}
    \mathbb{E}[\prod_{i=1}^N X_{i,t}^{k_i}|\mathcal{F}_{t-1}] = \prod_{i=1}^N \mathbb{E}[X_{i,t}^{k_i}|\mathcal{F}_{t-1}]
\end{equation}

For any sequence of vectors $\{v_1,\dots,v_n\}$, define $\text{outvec}(v_1,\dots,v_n) = \text{vec}(\dots\text{vec}(v_1 v_2^T)\dots v_3^T)$ to be a sequence of outer product and vectorizing operations. Let $\otimes$ be the Kronecker product. Fix $m \geq 0$, (\ref{eq:ngnar-cond_moment}) and (\ref{eq:ngnar-cond_indep}) together imply that 

\begin{equation}
    \mathbb{E}[\Tilde{\mathbf{X}}_t^{(m)} |\mathcal{F}_{t-1}] \leq^*  \Tilde{A}^{(m)} \Tilde{\mathbf{X}}_{t-1}^{(m)}  + \Tilde{v}^{(m)} 
\end{equation}

where $\Tilde{\mathbf{X}}_t^{(m)} = \text{outvec}(\mathbf{X}_t \times m) $, $\Tilde{v}^{(m)} $ is a constant vector, $\Tilde{A}^{(m)} $ is a block upper-triangular matrix whose blocks are of different sizes, 

\begin{equation}
    \Tilde{A}^{(m)} = \begin{pmatrix}
        \Tilde{A}_{m,m}& \Tilde{A}_{m,m-1}& \dots& \Tilde{A}_{m,1}\\
        0 & \Tilde{A}_{m-1,m-1} & \ddots & \vdots \\
        0 & 0 & \ddots & \vdots\\
        0 &0 & 0 & \Tilde{A}_{1,1}
    \end{pmatrix}
\end{equation}

with $\Tilde{A}_{k,k} = \underbrace{|A_1| \otimes |A_1| \dots \otimes |A_1|}_{k}$. The off-diagonal blocks are less important. (\ref{ngnar_stat}) implies that $\| A_1\|_\infty < 1$ which implies $\| \Tilde{A}_{k,k} \|_\infty < 1$. Hence we have $\rho(\Tilde{A}^{(m)}) = \max_k (\rho(\Tilde{A}_{k,k})) < 1$. Therefore, we have,

\begin{align}
    \mathbb{E}[\Tilde{\mathbf{X}}_t^{(m)} |\mathcal{F}_{t-l}] = (I+\sum_{h=0}^{l-1} (\Tilde{A}^{(m)})^h) v + (\Tilde{A}^{(m)})^l \Tilde{\mathbf{X}}_{t-l}^{(m)} \\
    \implies \mathbb{E}[\Tilde{\mathbf{X}}_t^{(m)}] = \lim_{l \rightarrow \infty} \mathbb{E}[\Tilde{\mathbf{X}}_t^{(m)} |\mathcal{F}_{t-l}] = (I - \Tilde{A}^{(m)})^{-1} v^{(m)}
\end{align}

i.e., all moments or cross-moments with order $\leq m$ exist.
\end{proof}

\begin{proof}[Proof of Proposition~\ref{prop:ngnarasym_ls}]
For ease of notation, we consider the case where $\alpha$ is global and $\alpha_0 = 0$. The functions below when acting on vectors are assumed to be entry-wise. For the estimation of a softplus NGNAR$(p,[s_1,s_p])$ process, we have
    \begin{equation}
        \hat{\mathbf{X}}_{t|t-1}(\mathbf{\theta}) = s(\sum_{j=1}^p A_j \mathbf{X_{t-j}}).
    \end{equation}

    The derivatives of $s(x)$ are 

    \begin{equation}
        s'(x) = (1+\exp(-x))^{-1} \in (0,1) 
    \end{equation}

    \begin{equation}
        s''(x) = (1+\exp(-x))^{-1}(1+\exp(x))^{-1} \in (0,1) 
    \end{equation}

    Then, the partial derivatives of $\hat{\mathbf{X}}_{t|t-1}(\mathbf{\theta}))$ are,

    \begin{equation}
        \frac{\partial \hat{\mathbf{X}}_{t|t-1}(\mathbf{\theta})) }{\partial \alpha_j} = \mathbf{X}_{t-j}*s'(\sum_{j=1}^p A_j \mathbf{X}_{t-j}) \leq^* X_{t-j} \quad a.s.
    \end{equation}

    \begin{equation}
        \frac{\partial \hat{\mathbf{X}}_{t|t-1}(\mathbf{\theta})) }{\partial \beta_{j,r}} = W^{(r)}\mathbf{X}_{t-j}*s'(\sum_{j=1}^p A_j \mathbf{X}_{t-j}) \leq^* W^{(r)}X_{t-j} \quad a.s. 
    \end{equation}

    \begin{equation}
        \frac{\partial^2 \hat{\mathbf{X}}_{t|t-1}(\mathbf{\theta})) }{\partial \alpha_{j_1} \alpha_{j_2}} = \mathbf{X}_{t-j_1}* \mathbf{X}_{t-j_2} *s''(\sum_{j=1}^p A_j \mathbf{X}_{t-j}) \leq^* \mathbf{X}_{t-j_1}* \mathbf{X}_{t-j_2} \quad a.s.
    \end{equation}

    \begin{equation}
        \frac{\partial^2 \hat{\mathbf{X}}_{t|t-1}(\mathbf{\theta})) }{\partial \alpha_{j_1} \partial \beta_{j_2,r}} = \mathbf{X}_{t-j_1}*(W^{(r)}X_{t-j_2}) *s''(\sum_{j=1}^p A_j \mathbf{X}_{t-j}) \leq^* \mathbf{X}_{t-j_1}*(W^{(r)}X_{t-j_2}) \quad a.s.
    \end{equation}

    \begin{equation}
        \frac{\partial^2 \hat{\mathbf{X}}_{t|t-1}(\mathbf{\theta})) }{\partial \beta_{j_1,r_1} \partial \beta_{j_2,r_2}} = (W^{(r_1)}X_{t-j_1})*(W^{(r_2)}X_{t-j_2}) *s''(\sum_{j=1}^p A_j \mathbf{X}_{t-j}) \leq^* (W^{(r_1)}X_{t-j_1})*(W^{(r_2)}X_{t-j_2}) \quad a.s.
    \end{equation}

    By Lemma~\ref{lem:ngnar-moment}, we have $\mathbb{E}[\|\frac{\partial \hat{\mathbf{X}}_{t|t-1}(\mathbf{\theta})) }{\partial \alpha_j}\|_2^2], \mathbb{E}[\| \frac{\partial \hat{\mathbf{X}}_{t|t-1}(\mathbf{\theta})) }{\partial \beta_{j,r}} \|_2^2] < \infty$, 
    
    $\mathbb{E}[\|\frac{\partial^2 \hat{\mathbf{X}}_{t|t-1}(\mathbf{\theta})) }{\partial \alpha_{j_1} \alpha_{j_2}}\|_2^2], \mathbb{E}[\|\frac{\partial^2 \hat{\mathbf{X}}_{t|t-1}(\mathbf{\theta})) }{\partial \alpha_{j_1} \partial \beta_{j_2,r}}\|_2^2], \mathbb{E}[\|\frac{\partial^2 \hat{\mathbf{X}}_{t|t-1}(\mathbf{\theta})) }{\partial \beta_{j_1,r_1} \partial \beta_{j_2,r_2}}\|_2^2] < \infty$ for all $j,r$.

    Let $d$ be the length of the $\theta$ vector.
Suppose there exists constants $a_1, \dots, a_d$ such that
\begin{equation}\label{eq:ngnar_ls_lin_indep}
    \mathbb{E}\left[\left\| \sum_{k=1}^d a_k \frac{\partial\hat{\mathbf{X}}_{t|t-1}(\mathbf{\theta})}{\partial\theta_k} \right\|^2\right] = 0
\end{equation}

    Since $s'(x) > 0$ for all $x$ and $s'(\sum_{j=1}^p A_j \mathbf{X}_{t-j})$ exists in all $\frac{\partial\hat{\mathbf{X}}_{t|t-1}(\mathbf{\theta})}{\partial\theta_k}$, for the same reason as the proof of Proposition~\ref{prop:gnariasym}, we have $a_1=\dots=a_d = 0$.

    The third-order derivatives are of the form

    \begin{equation}
        \frac{\partial^3 \hat{\mathbf{X}}_{t|t-1}(\mathbf{\theta})}{\partial\theta_{k_1} \partial\theta_{k_2} \partial\theta_{k_3}} = (W^{(r_1)}\mathbf{X}_{t-j_1})*(W^{(r_2)}\mathbf{X}_{t-j_2})*(W^{(r_3)}\mathbf{X}_{t-j_2})*s'''(\sum_{j=1}^p A_j \mathbf{X}_{t-j})
    \end{equation}

    where $r_1, r_2, r_3$ could be $0$ and $W^{(0)}$ is defined to be identity. We can check that $|s'''(x)| < 1$ for all $x$. Hence,

    \begin{equation}
        | \frac{\partial^3 \hat{\mathbf{X}}_{t|t-1}(\mathbf{\theta})}{\partial\theta_{k_1} \partial\theta_{k_2} \partial\theta_{k_3}} | \leq* (W^{(r_1)}\mathbf{X}_{t-j_1})*(W^{(r_2)}\mathbf{X}_{t-j_2})*(W^{(r_3)}\mathbf{X}_{t-j_2})
    \end{equation}

    Thus, we have that both 

    \begin{equation}
        \| \frac{\partial \hat{\mathbf{X}}_{t|t-1}(\mathbf{\theta})}{\partial\theta_{k_1}} * \frac{\partial^2 \hat{\mathbf{X}}_{t|t-1}(\mathbf{\theta})}{\partial\theta_{k_2} \partial\theta_{k_3}} \|^2, \leq G_{k_1,k_2,k_3}(\mathbf{X}_{t-1},\dots,\mathbf{X}_{t-p}) 
    \end{equation}

    \begin{equation}
        \| ({\mathbf{X}}_{t}- \hat{\mathbf{X}}_{t|t-1}(\mathbf{\theta}))*\frac{\partial^3 \hat{\mathbf{X}}_{t|t-1}(\mathbf{\theta})}{\partial\theta_{k_1} \partial\theta_{k_2} \partial\theta_{k_3}} \|^2 \leq H_{k_1,k_2,k_3}(\mathbf{X}_{t},\dots,\mathbf{X}_{t-p}) 
    \end{equation}

    where both $G_{k_1,k_2,k_3}(\mathbf{X}_{t-1},\dots,\mathbf{X}_{t-p}) $ and $ H_{k_1,k_2,k_3}(\mathbf{X}_{t},\dots,\mathbf{X}_{t-p})$ are polynomials of finite orders. By Lemma~\ref{lem:ngnar-moment} and Cauchy-Schwartz inequality, we have that $\mathbb{E}(G), \mathbb{E}(H) < \infty$. 

    Thus, by \cite{estnonlinear} Theorem~3.1, we have that $\hat{\theta}^{(n)} \xrightarrow{a.s} \theta^0$.

    Let $f_{t|t-1}(\theta)$ be (\ref{eq:f}), $R$ be (\ref{eq:R}). As we have assumed conditional independent Poisson distribution,

    \begin{equation}
        f_{t|t-1}(\theta) = \text{diag}(s(\sum_{j=1}^p A_j \mathbf{X}_{t-j}))
    \end{equation}

    Then using the fact that $s(x) \leq \log(2)+x$, $s'(x) < 1$, and lemma~\ref{lem:ngnar-moment}, we can show $R < \infty$ using the same strategy as in the proof of Proposition~\ref{prop:gnariasym}, which implies asymptotic normality by \cite{estnonlinear} Theorem~3.2.
    
\end{proof}

\begin{proof}[Proof of Proposition~\ref{prop:ngnarasym_mle}]

The first-order derivative of $l_n(\theta)$ w.r.t $\theta_k$ is

\begin{equation}
    \frac{\partial l_n(\theta)}{\partial \theta_k} = \sum_{t=p+1}^n \sum_{i=1}^N (\frac{X_{i,t}}{s(\Tilde{M}_{i,t})}-1) s'(\Tilde{M}_{i,t}) \frac{\partial \Tilde{M}_{i,t}}{\partial \theta_k}
\end{equation}

where $\Tilde{M}_{i,t} = g^{-1}(M_{i,t})$ which is the conditional mean before applying the response function, $\frac{\partial \Tilde{M}_{i,t}}{\partial \alpha_j} = X_{i,t-j}$, $\frac{\partial \Tilde{M}_{i,t}}{\partial \beta_{j,r}} = \sum_{q\in \mathcal{N}^r_t(i)} (w_{i,q}  X_{q,t-j})$ for $j=1,\dots,p$ and $r = 1,\dots, s_j$, $\frac{\partial \Tilde{M}_{i,t}}{\partial \alpha_0} = 1$. By construction, we know $\mathbb{E}[\frac{X_{i,t}}{s(\Tilde{M}_{i,t}(\theta^0))}-1|\mathcal{F}_{t-1}] = 0$. Hence,

\begin{equation}\label{eq:ngnarasym_mle_0mean}
    \mathbb{E}[(\frac{X_{i,t}}{s(\Tilde{M}_{i,t}(\theta^0))}-1) g'(\Tilde{M}_{i,t}) \frac{\partial \Tilde{M}_{i,t}(\theta^0)}{\partial \theta_k}] = \mathbb{E}[\mathbb{E}[(\frac{X_{i,t}}{s(\Tilde{M}_{i,t}(\theta^0))}-1) s'(\Tilde{M}_{i,t}(\theta^0)) \frac{\partial \Tilde{M}_{i,t}(\theta^0)}{\partial \theta_k}|\mathcal{F}_{t-1}]]= 0
\end{equation}

Moreover, 

\begin{align}
    &\mathbb{E}[((\frac{X_{i,t}}{s(\Tilde{M}_{i,t})}-1) s'(\Tilde{M}_{i,t}))^2|\mathcal{F}_{t-1}] = (\frac{s'(\Tilde{M}_{i,t})}{s(\Tilde{M}_{i,t})})^2 \leq 1 \quad \quad \text{a.s.}\\
    \implies &\mathbb{E}[((\frac{X_{i,t}}{s(\Tilde{M}_{i,t})}-1) s'(\Tilde{M}_{i,t}))^2] \leq 1
\end{align}

By Lemma~\ref{lem:ngnar-moment}, we also have $\mathbb{E}[|\frac{\partial \Tilde{M}_{i,t}}{\partial \theta_k}|^2] < \infty$. By Cauchy-Schwartz inequality, $\mathbb{E}[|(\frac{X_{i,t}}{\Tilde{M}_{i,t}}-1) s'(\Tilde{M}_{i,t}) \frac{\partial \Tilde{M}_{i,t}}{\partial \theta_k}|] < \infty$. Thus, the ergodicity of $\mathbf{X}_t$ guaranteed according to Lemma~\ref{lem:ngnar-sty} implies that 

\begin{equation}
    n^{-1} \frac{\partial l_n(\theta^0)}{\partial \theta_k} \xrightarrow{a.s.} 0
\end{equation}

The second-order derivatives of $l_n(\theta)$ are

\begin{equation}
    \frac{\partial^2 l_n(\theta) }{\partial \theta_{k_1} \partial \theta_{k_2}} = \sum_{t=p+1}^n \sum_{i=1}^N [(\frac{X_{i,t}}{s(\Tilde{M}_{i,t})}-1) s''(\Tilde{M}_{i,t})- \frac{X_{i,t}}{s^2(\Tilde{M}_{i,t})}(s'(\Tilde{M}_{i,t}))^2] \frac{\partial \Tilde{M}_{i,t}}{\partial \theta_{k_1}}\frac{\partial \Tilde{M}_{i,t}}{\partial \theta_{k_2}}
\end{equation}

Since $\frac{s''(x)}{s(x)}- (\frac{s'(x)}{s(x)})^2 < 0$ and $s''(x) > 0$ for all $x$, $(\frac{X_{i,t}}{s(\Tilde{M}_{i,t})}-1) s''(\Tilde{M}_{i,t})- \frac{X_{i,t}}{s^2(\Tilde{M}_{i,t})}(s'(\Tilde{M}_{i,t}))^2<0$ almost surely, which implies that $-\frac{\partial^2 l_n(\theta) }{\partial \theta^2 }$ is by definition semi-positive definite almost surely. Now, since $(\frac{s''(x)}{s(x)})^2 \leq 1$ for all $x$, using the same logic as above, we have 

\begin{equation}
    \mathbb{E}[|(\frac{X_{i,t}}{s(\Tilde{M}_{i,t})}-1) s''(\Tilde{M}_{i,t}) \frac{\partial \Tilde{M}_{i,t}}{\partial \theta_{k_1}}\frac{\partial \Tilde{M}_{i,t}}{\partial \theta_{k_2}}|] < \infty
\end{equation}

For the second part, we have by Lemma~\ref{lem:ngnar-moment},

\begin{equation}
    \mathbb{E}[(\frac{X_{i,t}}{s^2(\Tilde{M}_{i,t})}(s'(\Tilde{M}_{i,t}))^2)^2] = \mathbb{E}[s(\Tilde{M}_{i,t}) (s(\Tilde{M}_{i,t})+1) (\frac{s'(\Tilde{M}_{i,t})}{s(\Tilde{M}_{i,t})})^4] < \infty
\end{equation}

    Again, using Cauchy-Schwartz inequality and ergodic theorem, we obtain

    \begin{align}
    \begin{split}
        &- n^{-1} \frac{\partial^2 l_n(\theta) }{\partial \theta_{k_1} \partial \theta_{k_2}}\\
        \xrightarrow{a.s.} &\sum_{i=1}^N \mathbb{E}[ [\frac{X_{i,t}}{s^2(\Tilde{M}_{i,t})}(s'(\Tilde{M}_{i,t}))^2-(\frac{X_{i,t}}{s(\Tilde{M}_{i,t})}-1) s''(\Tilde{M}_{i,t})]\frac{\partial \Tilde{M}_{i,t}}{\partial \theta_{k_1}}\frac{\partial \Tilde{M}_{i,t}}{\partial \theta_{k_2}}]\\
        = &\sum_{i=1}^N \mathbb{E}[U]_{k_1,k_2}
    \end{split}
    \end{align}

    where $U$ is positive definite due to the same reason as (\ref{eq:ngnar_ls_lin_indep}) if we treat $U$ as a self-outer product.

    The third-order derivatives are

    \begin{align}
    \begin{split}
        &\frac{\partial^3 l_n(\theta)}{\partial \theta_{k_1} \partial \theta_{k_2} \partial \theta_{k_3}}\\ 
        = &\sum_{t=p+1}^n \sum_{i=1}^N \{[(\frac{X_{i,t}}{s(\Tilde{M}_{i,t})}-1) s'''(\Tilde{M}_{i,t}) - 3\frac{X_{i,t}}{s^2(\Tilde{M}_{i,t})} s'(\Tilde{M}_{i,t})s''(\Tilde{M}_{i,t}) + 2\frac{X_{i,t}}{s^3(\Tilde{M}_{i,t})}(s'(\Tilde{M}_{i,t}))^3]\\
        &\frac{\partial \Tilde{M}_{i,t}}{\partial \theta_{k_1}}\frac{\partial \Tilde{M}_{i,t}}{\partial \theta_{k_2}}\frac{\partial \Tilde{M}_{i,t}}{\partial \theta_{k_3}}\} = \sum_{t=p+1}^n \sum_{i=1}^N Z_{i,t}(\theta, {k_1},{k_2},{k_3})
    \end{split}
    \end{align}

    We can show that $|\frac{s'''(x)}{s(x)}|, |\frac{s'(x)s''(x)}{s^2(x)}|, |\frac{s'(x)}{s(x)}| < 1$. Hence, we have

    \begin{equation}
        |Z_{i,t}(\theta, {k_1},{k_2},{k_3})| \leq (C |X_{i,t}| + D)|\frac{\partial \Tilde{M}_{i,t}}{\partial \theta_{k_1}}\frac{\partial \Tilde{M}_{i,t}}{\partial \theta_{k_2}}\frac{\partial \Tilde{M}_{i,t}}{\partial \theta_{k_3}}|\} 
    \end{equation}

    where $C,D > 0$ are some fixed constant uniformly in $t,i,k_1,k_2,k_3$ and $\theta$. By Lemma~\ref{lem:ngnar-moment} and Cauchy-Schwartz inequality, $\mathbb{E}[(C |X_{i,t}| + D)|\frac{\partial \Tilde{M}_{i,t}}{\partial \theta_{k_1}}\frac{\partial \Tilde{M}_{i,t}}{\partial \theta_{k_2}}\frac{\partial \Tilde{M}_{i,t}}{\partial \theta_{k_3}}|] < \infty$. Using ergodicity, we have

    \begin{equation}
        n^{-1}|\frac{\partial^3 l_n(\theta)}{\partial \theta_{k_1} \partial \theta_{k_2} \partial \theta_{k_3}}| \xrightarrow{a.s} \mathbb{E}[|\sum_{i=1}^N Z_{i,t}(\theta, {k_1},{k_2},{k_3})|] < \infty
    \end{equation}

    \cite{estnonlinear}~Theorem 2.1 thus implies that $\hat{\theta}_M^{(n)} \xrightarrow{a.s} \theta^0$.

    Let $a_1, \dots, a_d$ be an arbitrary sequence of real numbers. Let $S_{i,t,k}(\theta) = \sum_{i=1}^N(\frac{X_{i,t}}{s(\Tilde{M}_{i,t}(\theta))}-1) s'(\Tilde{M}_{i,t}(\theta)) \frac{\partial \Tilde{M}_{i,t}(\theta)}{\partial \theta_k}$. From (\ref{eq:ngnarasym_mle_0mean}), we know that the increments of $\sum_{k=1}^d a_k \frac{\partial l_n(\theta^0)}{\partial \theta_k}$ satisfy

    \begin{equation}
        \mathbb{E}[\sum_{k=1}^da_kS_{i,t,k}(\theta^0)|\mathcal{F}_{t-1}] = 0
    \end{equation}

    Then we know that $\sum_{k=1}^d a_k \frac{\partial l_n(\theta^0)}{\partial \theta_k}$ is a strictly stationary ergodic martingale process. Similar to above, we can show that the second moment of the increment $\sigma^2_t$ is always finite. Thus, by \cite{BILLINGSLEY1961},

    \begin{equation}\label{eq:ngnar_mle_lin_normal}
        n^{-1/2}\sum_{k=1}^d a_k \frac{\partial l_n(\theta)}{\partial \theta_k} \xrightarrow{D} N(0,\sigma^2_t)
    \end{equation}

    Now, Let $R$ be the matrix such that $R_{k_1,k_2} = \mathbb{E}[S_{i,t,k_1}(\theta^0)S_{i,t,k_2}(\theta^0)]$. We can show that $\mathbb{E}[|S_{i,t,k_1}(\theta^0)S_{i,t,k_2}(\theta^0)|] < \infty$ using the same way as before. Then by ergodic theorem, 
    
    \begin{equation}\label{eq:ngnar_mle_cov}
        n^{-1} \frac{\partial l_n(\theta)}{\partial \theta_{k_1}}\frac{\partial l_n(\theta)}{\partial \theta_{k_2}} \xrightarrow{a.s.} R_{k_1,k_2}
    \end{equation}
    
    which implies in conjuncture with (\ref{eq:ngnar_mle_lin_normal}) that $n^{-1/2} \frac{\partial l_n(\theta)}{\partial \theta} \xrightarrow{D} MVN(0,R)$. The result thus follows from \cite{estnonlinear}~Theorem 2.2. 
\end{proof}

\bibliographystyle{guy3}
\bibliography{ref-2}

\end{document}